\documentclass[final,3p,times,nopreprintline,numafflabel]{elsarticle}
\usepackage{makecell}
\usepackage{ragged2e}
\usepackage{caption}
\usepackage{overpic}
\usepackage{xcolor}
\usepackage{booktabs}
\usepackage{tabularx}
\usepackage{longtable}
\usepackage{amssymb}
\usepackage{amsthm}
\usepackage{amsmath}
\usepackage{multirow}
\usepackage{listings}
\usepackage{placeins}
\usepackage{needspace}
\usepackage{url}
\usepackage[hidelinks]{hyperref}
\begin{document}
	
	\begin{frontmatter}
		
\title{ZStar: A unified toolkit for polarization, Born effective charges, dielectric and piezoelectric responses, and infrared and Raman spectra}
		
		\author[a]{Xudong Zhu\corref{author1}}
		\author[a,b,c]{Junhong Li}
		\author[a]{Gan Jin}
		\author[a]{Zheng Guan}
		\author[a]{Meng Zhang}
		\author[a,b,c]{Lixin He \corref{author2}}
		
		\cortext[author1]{Corresponding author. \textit{E-mail address:} zhu@iai.ustc.edu.cn}
		\cortext[author2]{Corresponding author. \textit{E-mail address:} helx@ustc.edu.cn}
		
		\address[a]{Institute of Artificial Intelligence, Hefei Comprehensive National Science Center, Hefei, Anhui, 230088, China}
		\address[b]{Laboratory of Quantum Information, University of Science and Technology of China, Hefei, Anhui, 230026, China}
		\address[c]{Hefei National Laboratory, University of Science and Technology of China, Hefei, Anhui, 230088, China}
		
		\begin{abstract}
Polarization and the response to atomic displacements determine the dielectric, piezoelectric, and vibrational spectroscopy of materials, yet these quantities are usually obtained through separate workflows within ABACUS. Here, we present ZStar, an open-source Python toolkit that unifies these calculations. Using ABACUS and PYATB, ZStar derives Born effective charges (BECs) and zone-center force constants from one symmetry-adapted displacement set, and reuses the displaced electronic matrices for Raman spectra. Berry phases in periodic directions combined with charge-density dipoles in nonperiodic directions extend the method from bulk crystals (3D) to 2D and 1D materials and molecules (0D). These BECs and phonon modes give infrared intensities and the lattice contribution to the dielectric response. From finite strains, ZStar further obtains clamped-ion and relaxed-ion piezoelectric coefficients. This unified workflow reduces the computational cost by up to 4.2$\times$ for BECs and zone-center phonons and 8.4$\times$ for combined infrared and Raman calculations. ZStar also supports machine-learned models by generating training data and post-processing prediction results into physical properties.
		\end{abstract}

		
	\end{frontmatter}
	
{\bf PROGRAM SUMMARY}

\begin{small}
	\noindent
	{\em Program Title: ZStar}                                          \\
	{\em Python package:} \url{https://pypi.org/project/zstar/} \\
	{\em Developer's repository link: \url{https://github.com/xdzhu/ZStar}} \\
	{\em Licensing provisions: GNU General Public License 3 (GPLv3)} \\
	{\em Programming language: Python} \\
	{\em Nature of problem: Separate BEC and phonon calculations can repeat the same electronic-structure calculations on displaced cells. For molecules, wires, and two-dimensional materials, the electrical response also depends on the periodic and open directions.} \\
	{\em Solution method: ZStar combines symmetry-adapted displacement and strain responses to obtain BEC, zone-center force constants, and bulk piezoelectric coefficients and to automate dielectric and vibrational analysis. Phase-wrapped branch matching and real-space dipoles treat periodic and nonperiodic directions, respectively, enabling calculations for bulk crystals, two-dimensional materials, one-dimensional structures, and molecules.} \\
\end{small}

\clearpage

\section{Introduction}

Polarization and dielectric response connect atomic-scale charge redistribution to the macroscopic behavior of ferroelectric and piezoelectric materials \cite{Rabe2007Ferroelectrics,Spaldin2012}. The Born effective charge (BEC) tensor describes how polarization changes as an atom is displaced. BECs enter the dipole interaction responsible for longitudinal--transverse optical (LO--TO) splitting and, together with the phonon modes, determine infrared (IR) intensities and the lattice dielectric response \cite{Ghosez1995,Gonze1997,Baroni2001}. Raman intensities depend on how the electronic polarizability changes along a vibrational mode \cite{Porezag1996,Lazzeri2003,Veithen2005}. Calculating these responses from first principles helps interpret measurements of ferroelectric switching \cite{Wang2026}, electromechanical coupling \cite{Kim2015AlNPiezo}, and vibrational spectra \cite{Lee2010MoS2}.

Born effective charges and force constants can be obtained using density-functional perturbation theory (DFPT) or finite displacement method. Separate finite-displacement workflows use different sets of structures to evaluate polarization changes for BECs and forces for phonons. Direct calculations of Raman tensors add displacements along the normal modes. Each set requires its own self-consistent calculations. For a given displaced structure, one calculation can provide forces and the electronic data needed for polarization and optical response. For low-dimensional materials such as ferroelectric monolayer $\alpha$-In$_2$Se$_3$ \cite{Ding2017,Zhou2017,Xiao2018}, the workflow must also accommodate both periodic and open directions.

ASE provides tools for phonons and molecular vibrations \cite{Larsen2017ASE}, atomate2 organizes materials simulations \cite{Ganose2025Atomate2}, and AiiDA-Vibroscopy provides workflows for vibrational spectroscopy \cite{BastoneroMarzari2024}. Combining several response calculations requires a common reference structure and consistent conventions for atomic displacements and tensor components. A unified toolkit bringing these response calculations together is needed for ABACUS \cite{Li2016,ABACUS2025}.

Here, we present ZStar, an open-source toolkit that obtains BECs and zone-center force constants from polarization and forces evaluated for the same symmetry-adapted displacements. It reuses the displaced electronic matrices to calculate nonresonant Raman tensors. Phonopy generates the displacements \cite{Togo2015,Togo2023}, ABACUS supplies forces and matrices, and PYATB evaluates the Berry-phase polarization and electronic dielectric tensor \cite{Jin2023}. ZStar combines Berry-phase polarization in periodic directions \cite{KingSmith1993,Vanderbilt2018Book} with charge-density dipoles in open directions, covering bulk crystals, two-dimensional materials, nanowires, and molecules. A finite-strain workflow yields clamped-ion and relaxed-ion piezoelectric coefficients for insulating bulk crystals \cite{Vanderbilt2000,Wu2005}. ZStar also generates first-principles training data and post-processes predictions from machine-learned interatomic potentials and Hamiltonian models to obtain material properties. Table~\ref{table:zstar_capabilities} lists the main capabilities.

\begin{table*}[h]
	\centering
\small
	\caption{Main capabilities of ZStar}
\begin{tabularx}{0.92\textwidth}{>{\centering\arraybackslash}p{0.25\textwidth}X}
		\hline
		Module & Functions \\
		\hline
\multirow{3}{*}{\mbox{Polarization}}
		& Berry-phase polarization and real-space dipoles \\
		& symmetry-adapted finite-displacement BECs \\
		& polarization branch matching and acoustic sum rules \\
		\hline
\multirow{3}{*}{\mbox{Phonon}}
		& zone-center force constants and phonon irreducible representations \\
		& mode-resolved phonon analysis \\
		& finite-wavevector phonon dispersion, DOS, and optional bulk NAC \\
		\hline
\multirow{2}{*}{\mbox{Spectroscopy}}
		& infrared and Raman activity classification \\
		& infrared and Raman spectra \\
		\hline
\multirow{2}{*}{Dielectric}
		& electronic and phonon dielectric response \\
		& frequency-dependent dielectric spectra \\
		\hline
\multirow{2}{*}{\mbox{Piezoelectric}}
		& proper bulk piezoelectric tensors and internal-relaxation contributions \\
		& elastic response and conversion from stress to strain coefficients \\
		\hline
	\end{tabularx}
	\label{table:zstar_capabilities}
\end{table*}
\vspace{-5mm}

\section{Theory}
\label{sec:theory}

\begin{figure*}[htbp]
	\begin{center}
		\includegraphics[width=1.0\textwidth]{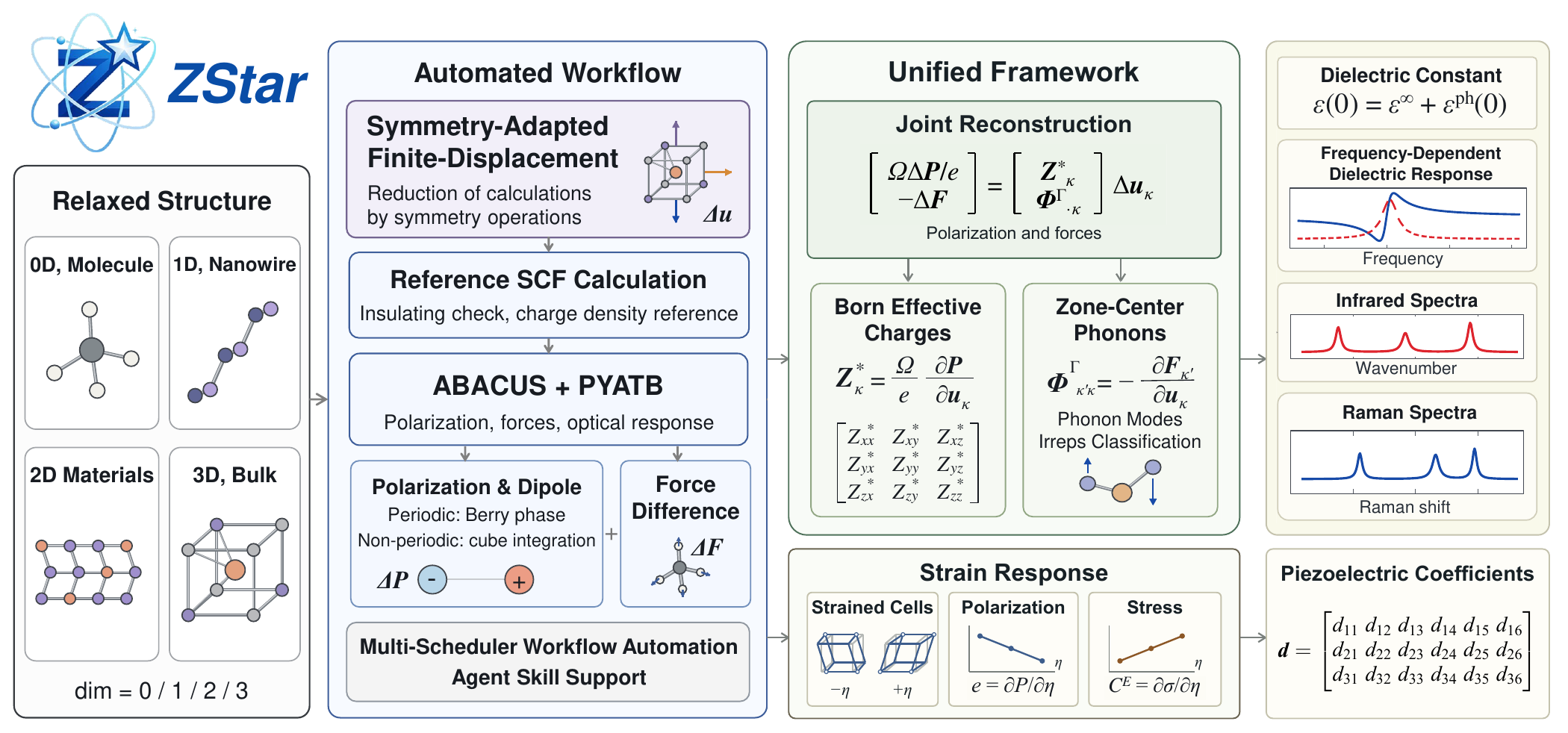}
		\caption{ZStar workflow with ABACUS and PYATB. The linear reconstruction uses the polarization and forces from the same symmetry-adapted atomic displacements to obtain BECs and zone-center force constants. ZStar reuses the displaced-cell Hamiltonian, overlap, and position matrices to calculate the electronic dielectric derivatives for Raman spectra. The finite-strain workflow gives the piezoelectric and elastic tensors from polarization and stress.}
		
		\label{fig:zstar_workflow}
	\end{center}
\end{figure*}

Figure~\ref{fig:zstar_workflow} shows how ZStar obtains material responses from atomic displacements and homogeneous strains. For the displacement workflow, ZStar reconstructs BECs and zone-center force constants jointly from the polarization and forces. The resulting BECs and phonon modes give the infrared intensities and the phonon contribution to the dielectric response. The same displaced electronic matrices provide the dielectric derivatives for Raman scattering. The finite-strain workflow obtains the piezoelectric and elastic tensors from the polarization and stress, respectively.

\vspace{-8mm}

\subsection{Polarization and Born effective charge}

For a periodic insulator, the modern theory expresses the macroscopic polarization as an ionic term plus an electronic Berry-phase term \cite{KingSmith1993,Resta1994,Vanderbilt2018Book},
\begin{equation}
 \boldsymbol{P}
 =\frac{e}{\Omega}\sum_{\kappa}Z_{\kappa}^{\mathrm{ion}}\boldsymbol{r}_{\kappa}
 -\frac{e}{(2\pi)^3}\sum_n f_n\int_{\mathrm{BZ}}
 \boldsymbol{\mathcal A}_n(\boldsymbol{k})\,\mathrm{d}^3k,
 \qquad
 \boldsymbol{\mathcal A}_n(\boldsymbol{k})
 =i\langle u_{n\boldsymbol{k}}|\nabla_{\boldsymbol{k}}u_{n\boldsymbol{k}}\rangle .
 \label{eq:berry_phase_polarization}
\end{equation}
Here $Z_{\kappa}^{\mathrm{ion}}$ and $\boldsymbol{r}_{\kappa}$ are the ionic valence and position, $f_n$ is the band occupation, and $\boldsymbol{\mathcal A}_n$ is the Berry connection of the cell-periodic state $u_{n\boldsymbol{k}}$. Equation~(\ref{eq:berry_phase_polarization}) is defined modulo a polarization quantum, so only branch-consistent polarization changes are unique. The BEC tensor is the linear derivative of this polarization with respect to an atomic displacement. For atom $\kappa$, it is \cite{Gonze1997,Wu2005}
\begin{equation}
	Z^{*}_{\kappa,\alpha\beta}
	=
	\frac{\Omega}{e}
	\frac{\partial P_{\alpha}}{\partial u_{\kappa\beta}},
	\label{eq:bec_definition}
\end{equation}
where $\Omega$ is the cell volume, $e$ is the elementary charge, $P_{\alpha}$ is the polarization component along Cartesian direction $\alpha$, and $u_{\kappa\beta}$ is the displacement of atom $\kappa$ along Cartesian direction $\beta$. The factor $1/e$ makes $Z^*$ dimensionless. Its entries give the effective charges in units of $e$.

ZStar evaluates $P_{\alpha}$ from the Berry phase in periodic directions. For an isolated molecule, it differentiates the molecular dipole moment.

\subsection{Unified framework for Born effective charges and zone-center phonons}
\label{sec:unified_response}

\begin{figure*}[!htbp]
 \centering
 \includegraphics[width=0.9\textwidth]{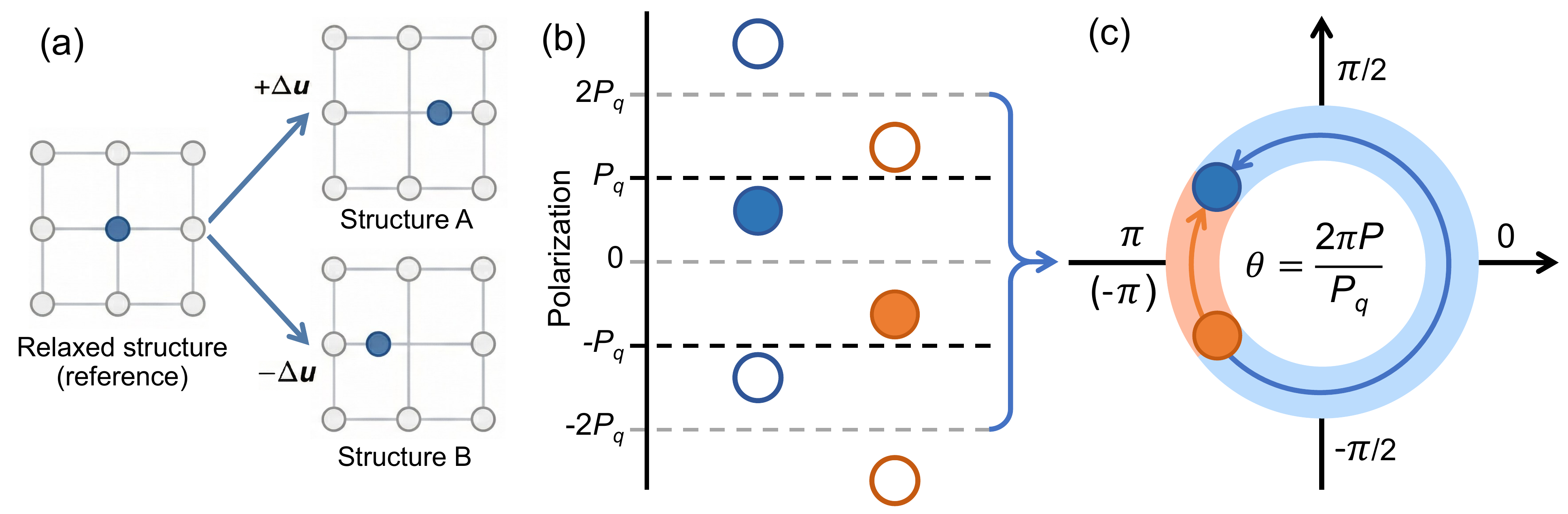}
 \caption{Local finite-displacement polarization response and phase-wrapped branch matching. (a) Reference and nearby displaced structures supply the polarization changes used in BEC evaluation. (b) Equivalent branches along a polarization-quantum direction are separated by $P_q$; local matching removes the integer branch offset. (c) The same matching is represented by the shortest signed arc on the phase circle, with $\theta=2\pi P/P_q$. The matched changes are transformed to Cartesian coordinates before response-tensor reconstruction.}
 \label{fig:fd_branch_matching}
\end{figure*}

Finite-displacement calculations determine BECs from polarization changes and harmonic force constants from force changes. For opposite displacements $\pm\Delta u_{\kappa\beta}$ of atom $\kappa$ along Cartesian direction $\beta$, central differences give
\begin{equation}
 Z^*_{\kappa,\alpha\beta}
 \simeq\frac{\Omega}{e}\,
 \frac{P_\alpha(+\Delta u_{\kappa\beta})-P_\alpha(-\Delta u_{\kappa\beta})}
      {2\Delta u_{\kappa\beta}},
 \label{eq:cartesian_response_differences}
\end{equation}
\begin{equation}
 \Phi^\Gamma_{\kappa'\alpha,\kappa\beta}
 \simeq-
 \frac{F_{\kappa'\alpha}(+\Delta u_{\kappa\beta})-F_{\kappa'\alpha}(-\Delta u_{\kappa\beta})}
      {2\Delta u_{\kappa\beta}} .
 \label{eq:cartesian_force_differences}
\end{equation}
The polarization difference is evaluated on a locally continuous branch, and $F_{\kappa'\alpha}$ is the force on atom $\kappa'$ along $\alpha$. Here $2\Delta u_{\kappa\beta}$ is the change in the displaced atom's coordinate along the displacement axis, read from the positive and negative displacement structures. A forward difference uses the corresponding coordinate difference between the positive displacement and the reference. Sampling $\beta=x,y,z$ supplies the three columns of each tensor; the paired finite-displacement construction is shown schematically in Fig.~\ref{fig:fd_branch_matching}(a).
ZStar extracts both responses from each pair of calculations and uses site symmetry to select the displacement directions.

At fixed lattice vectors and zero applied electric field, BECs and force constants follow from mixed derivatives of the electric enthalpy \cite{Wu2005}. For a small displacement $\boldsymbol{d}$ of atom $\kappa$ in an insulator, the polarization and force expansions about the reference structure are
\begin{equation}
 \frac{\Omega\Delta\boldsymbol{P}}{e}
  =Z^*_\kappa\boldsymbol{d}+O(\lVert\boldsymbol{d}\rVert^2),
 \label{eq:unified_observations}
\end{equation}
\begin{equation}
 -\Delta\boldsymbol{F}_{\kappa'}
  =\Phi^\Gamma_{\kappa'\kappa}\boldsymbol{d}+O(\lVert\boldsymbol{d}\rVert^2).
 \label{eq:unified_force_observations}
\end{equation}
For opposite displacements, the even terms cancel and the central-difference derivative has an error of order $\lVert\boldsymbol d\rVert^2$. The force change is $\Delta\boldsymbol F_{\kappa'}=\boldsymbol F_{\kappa'}(\boldsymbol d)-\boldsymbol F_{\kappa'}(\boldsymbol 0)$. Along an open direction, ZStar uses the dipole change in place of the corresponding component of $\Omega\Delta\boldsymbol P$.

Displacing an atom in the input cell also displaces its periodic images. The resulting matrix is consequently the zone-center, lattice-summed force constant,
\begin{equation}
 \Phi^\Gamma_{\kappa'\kappa}
 =\sum_{\boldsymbol{R}}\Phi_{\kappa'\kappa}(\boldsymbol{R}).
 \label{eq:gamma_force_constants}
\end{equation}
The matrix determines the zone-center modes.

\subsection{Symmetry reduction of finite-displacement calculations}
\label{sec:symmetry_reduction}

ZStar displaces representative atoms along symmetry-inequivalent directions. The reconstruction uses operations that preserve the reference Hamiltonian and electrical boundary conditions.

Let $g$ have Cartesian rotation $R_g$ and atomic permutation $p_g$. Because displacement, polarization change, and force are polar vectors, the response tensors obey
\begin{equation}
 Z^*_{p_g(\kappa)}=R_gZ^*_\kappa R_g^{\mathrm T},
 \qquad
 \Phi^\Gamma_{p_g(\kappa'),p_g(\kappa)}
 =R_g\Phi^\Gamma_{\kappa'\kappa}R_g^{\mathrm T}.
 \label{eq:symmetry_reconstruction}
\end{equation}
For the site-symmetry operations satisfying $p_g(\kappa)=\kappa$, one computed displacement supplies the additional observations
\begin{equation}
 \boldsymbol{d}'=R_g\boldsymbol{d},\qquad
 \Delta\boldsymbol{P}'=R_g\Delta\boldsymbol{P},\qquad
 \Delta\boldsymbol{F}'_{p_g(\kappa')}=R_g\Delta\boldsymbol{F}_{\kappa'} .
 \label{eq:response_symmetry_images}
\end{equation}
Each rotation acts on the force directions and permutes the atoms that receive the forces.

Following the symmetry-based finite-displacement construction used in Phonopy \cite{Togo2015,Togo2023}, let $s=1,\ldots,M$ index the explicit and symmetry-generated observations for representative atom $\kappa$. For a cell containing $N$ atoms, each displaced structure provides one three-component polarization response and $N$ three-component force responses. These $N+1$ response blocks are collected in Eq.~(\ref{eq:response_blocks}) as
\begin{equation}
 \boldsymbol{y}^{(s)}_\kappa=
 \begin{bmatrix}
  \Omega\Delta\boldsymbol{P}^{(s)}/e\\
  -\Delta\boldsymbol{F}^{(s)}_1\\
  \vdots\\
  -\Delta\boldsymbol{F}^{(s)}_N
 \end{bmatrix}
 \in\mathbb{R}^{3(N+1)},
 \qquad
 B_\kappa=
 \begin{bmatrix}
  Z^*_\kappa\\
  \Phi^\Gamma_{1\kappa}\\
  \vdots\\
  \Phi^\Gamma_{N\kappa}
 \end{bmatrix}
 \in\mathbb{R}^{3(N+1)\times3}.
\label{eq:response_blocks}
\end{equation}
so that $\boldsymbol{y}^{(s)}_\kappa\simeq B_\kappa\boldsymbol{d}^{(s)}_\kappa$, with $\boldsymbol{d}^{(s)}_\kappa\in\mathbb{R}^3$. Real-space dipole changes replace the corresponding components of $\Omega\Delta\boldsymbol{P}^{(s)}$ along open directions. Collecting the observations as columns gives the matrices in Eq.~(\ref{eq:response_matrices}):
\begin{equation}
 X_\kappa=
 \begin{bmatrix}
  \boldsymbol{d}^{(1)}_\kappa&\cdots&\boldsymbol{d}^{(M)}_\kappa
 \end{bmatrix}
 \in\mathbb{R}^{3\times M},
 \qquad
 Y_\kappa=
 \begin{bmatrix}
  \boldsymbol{y}^{(1)}_\kappa&\cdots&\boldsymbol{y}^{(M)}_\kappa
 \end{bmatrix}
 \in\mathbb{R}^{3(N+1)\times M}.
\label{eq:response_matrices}
\end{equation}
ZStar solves the symmetry-expanded linear equations by least squares:
\begin{equation}
 Y_\kappa\simeq B_\kappa X_\kappa,\qquad
 B_\kappa=Y_\kappa X_\kappa^+,\qquad
 \operatorname{rank}X_\kappa=3,
 \label{eq:unified_rank}
\end{equation}
where $X_\kappa^+\in\mathbb{R}^{M\times3}$ is the Moore--Penrose pseudoinverse. The first three rows of $B_\kappa$ give $Z^*_\kappa$, and the following $N$ three-row blocks give $\Phi^\Gamma_{\kappa'\kappa}$ for $\kappa'=1,\ldots,N$. The full rank of the symmetry-expanded displacement matrix $X_\kappa$ ensures a unique linear map for each response block. Its conditioning controls the amplification of numerical errors. A low-symmetry site generally needs three independent seed directions. Fewer suffice when their symmetry images span Cartesian space.

For Raman spectra, each of these displacement structures also supplies $\Delta\boldsymbol{\epsilon}^{\infty,(s)}$. This adds the optional response block $\operatorname{vec}(\Delta\boldsymbol{\epsilon}^{\infty,(s)})$ to the observations in Eq.~(\ref{eq:response_blocks}). The same displacements and site-symmetry operations determine the dielectric derivatives.

Opposite displacements may be computed explicitly or supplied by a site-symmetry operation. When both signs are present with equal weights, their even-order response terms cancel, leaving a derivative with second-order truncation error in the displacement amplitude. When site symmetry cannot generate the opposite direction, the calculation includes the negative displacement. The fitted representative tensors are then transformed to equivalent atoms using Eq.~(\ref{eq:symmetry_reconstruction}), averaging equivalent symmetry images.

For a neutral system, invariance under a rigid translation imposes the BEC charge-neutrality condition. The force constants additionally satisfy translational invariance and reciprocity:
\begin{equation}
 \sum_\kappa Z^*_\kappa=0,\qquad
 \sum_\kappa\Phi^\Gamma_{\kappa'\kappa}=0,\qquad
 \Phi^\Gamma_{\kappa'\kappa}
 =\bigl(\Phi^\Gamma_{\kappa\kappa'}\bigr)^{\mathrm T}.
 \label{eq:asr}
\end{equation}
The BEC correction subtracts the full-cell mean from each atomic tensor,
\begin{equation}
 Z^*_\kappa
 =Z^{*,\mathrm{raw}}_\kappa
 -\frac{1}{N}\sum_{\kappa'}Z^{*,\mathrm{raw}}_{\kappa'}.
 \label{eq:born_projection}
\end{equation}
ZStar removes the components along rigid translations from the zone-center force-constant matrix to satisfy the acoustic sum rule. Let $T$ contain three orthonormal rigid-translation vectors and let $\Pi=I-TT^{\mathrm T}$. The projected matrix is
\begin{equation}
 \Phi^\Gamma
 =\Pi\,\frac{\Phi^{\Gamma,\mathrm{raw}}
             +(\Phi^{\Gamma,\mathrm{raw}})^{\mathrm T}}{2}\,\Pi .
 \label{eq:hessian_projection}
\end{equation}

\subsection{Phase-wrapped branch matching for multi-valued polarization}
\label{sec:branch_matching}
Because Berry-phase polarization is defined modulo a lattice of polarization quanta, the differences in Eq.~(\ref{eq:unified_observations}) are evaluated on a locally continuous branch \cite{Spaldin2012,KingSmith1993,Resta1994}. Let $\boldsymbol{q}_j=e\boldsymbol{a}_j/\Omega$ denote the polarization quantum associated with lattice vector $\boldsymbol{a}_j$.
\begin{equation}
 \boldsymbol{P}=\sum_{j=1}^{3}\xi_j\boldsymbol{q}_j,\qquad
 \xi_j\equiv\xi_j+n_j,\quad n_j\in\mathbb Z .
 \label{eq:polarization_quantum}
\end{equation}
Here $\xi_j$ denotes a dimensionless coefficient along polarization quantum $\boldsymbol{q}_j$. Reported values for nearby structures can therefore differ by an integer branch shift in addition to the physical response, as illustrated in Fig.~\ref{fig:fd_branch_matching}(b).

For two configurations $A$ and $B$ with the same cell, local branch matching selects integers that make each reduced difference $\Delta\xi_j=\xi_j^{(A)}-\xi_j^{(B)}-n_j$ continuous with the reference. For sufficiently small perturbations with $|\Delta\xi_j|<1/2$, this is the nearest branch in each reduced coordinate. Define $\theta_j=2\pi\xi_j$; the matched change is
\begin{equation}
 \Delta\xi_j
 =\frac{1}{2\pi}\operatorname{atan2}
 \left[\sin\!\left(\theta_j^{(A)}-\theta_j^{(B)}\right),
       \cos\!\left(\theta_j^{(A)}-\theta_j^{(B)}\right)\right],
 \qquad
 \Delta\boldsymbol{P}=\sum_j\Delta\xi_j\boldsymbol{q}_j .
 \label{eq:wrapped_deltaP}
\end{equation}
The wrapped phase is the orange arc plotted in Fig.~\ref{fig:fd_branch_matching}(c). ZStar converts each matched change from the polarization-quantum basis to Cartesian coordinates, including for nonorthogonal cells. This gives a local derivative along perturbations that keep the system insulating \cite{Vanderbilt2000,Bonini2020,Poteshman2026}. As for open directions, ZStar uses differences in the real-space dipole.

\subsection{Hybrid two-dimensional polarization and Born effective charges}
For a slab periodic in the $xy$ plane with vacuum along $z$, ZStar uses Berry-phase polarization in the plane and a real-space dipole along the slab normal. Dielectric screening determines the long-wavelength polar phonon dispersion in two dimensions \cite{Sohier2017}. ZStar checks that the reference structure is insulating before evaluating its in-plane derivatives. The in-plane BEC follows from the polarization per unit volume $\boldsymbol{P}^{3\mathrm{D}}$ calculated by PYATB:
\begin{equation}
 Z^{*}_{\kappa,\alpha\beta}
 =
 \frac{\Omega}{e}
 \frac{\partial P^{3\mathrm{D}}_{\alpha}}
 {\partial u_{\kappa\beta}},
 \qquad \alpha\in\{x,y\},
 \label{eq:2d_inplane_bec}
\end{equation}
and $\Omega=A L_z$ exactly compensates the $1/L_z$ dilution of the polarization per unit volume.

Along the open direction, the position operator is well defined for the localized, neutral slab charge distribution. The response can therefore be obtained from the total real-space dipole evaluated from a charge-density cube,
\begin{equation}
 p_z
 =
 e\sum_{\kappa} Z_{\kappa}^{\mathrm{ion}}z_{\kappa}
 -
 e\int_{\Omega}z\,n(\boldsymbol{r})\,\mathrm{d}^{3}r ,
 \label{eq:slab_dipole}
\end{equation}
where $Z_{\kappa}^{\mathrm{ion}}$ is the ionic valence charge and $n(\boldsymbol{r})$ is the electron number density. ZStar unwraps the coordinates and density on a common grid. For an isolated layer, the Berry phase and density integration yield the same dipole change, $\Omega\Delta P_z=\Delta p_z$ \cite{KingSmith1993,VanderbiltKingSmith1993}. Density integration reduces the post-processing time for hBN (Section~\ref{sec:two_dimensional_bec_results}). The out-of-plane BEC is
\begin{equation}
 Z^{*}_{\kappa,z\beta}
 =
 \frac{1}{e}
 \frac{\partial p_z}{\partial u_{\kappa\beta}} .
 \label{eq:2d_outofplane_bec}
\end{equation}
Equations~(\ref{eq:2d_inplane_bec}) and (\ref{eq:2d_outofplane_bec}) give the in-plane and normal polarization rows for all displacement directions in Eq.~(\ref{eq:bec_definition}). ZStar reconstructs the tensor using the symmetry operations of the two-dimensional slab and applies the constraints in Section~\ref{sec:symmetry_reduction}.

\subsection{Hybrid one-dimensional polarization and line polarizability}
\label{sec:one_dimensional}

For a neutral nanowire along $z$, ZStar calculates the axial polarization $P_z$ from the Berry phase and obtains the two transverse components by integrating the charge density,
\begin{equation}
 p_{\alpha}
 =
 e\sum_{\kappa}Z_{\kappa}^{\mathrm{ion}}r_{\kappa\alpha}
 -e\int_{\Omega}r_{\alpha}n(\boldsymbol{r})\,\mathrm d^3r,
 \qquad \alpha\in\{x,y\}.
 \label{eq:wire_transverse_dipole}
\end{equation}
The hybrid BEC is assembled as
\begin{equation}
 Z^{*}_{\kappa,\alpha\beta}
 =
 \begin{cases}
 \displaystyle\frac{1}{e}\,\frac{\partial p_{\alpha}}{\partial u_{\kappa\beta}},
 & \alpha\in\{x,y\},\\[2pt]
 \displaystyle\frac{\Omega}{e}\,\frac{\partial P_z^{3\mathrm D}}{\partial u_{\kappa\beta}},
 & \alpha=z.
 \end{cases}
 \label{eq:1d_hybrid_bec}
\end{equation}
ZStar unwraps the ionic coordinates and density grid about the valence-charge-weighted circular center of the reference nanowire. Using that center for all displacements keeps the vacuum cut outside the localized charge, including when the wire crosses a cell boundary.
ZStar multiplies the susceptibility per unit volume by the transverse cell area to obtain the electronic line polarizability
\begin{equation}
 \boldsymbol{\alpha}^{1\mathrm D}_{\mathrm{el}}
 =
 \frac{A_{\perp}}{4\pi}
 \left(\boldsymbol{\epsilon}^{\infty}_{\mathrm{sc}}-\boldsymbol{I}\right),
 \label{eq:1d_line_polarizability}
\end{equation}
in Gaussian area units, where $A_{\perp}$ is the cross-sectional area of the simulation cell. The line-response quantities defined below use the alternative SI-derived normalization $\alpha/\varepsilon_0$, also in area units; the two conventions differ by $4\pi$.

\subsection{Molecular polarization and atomic polar tensors}

For an isolated molecule in a non-periodic supercell, ZStar differentiates the molecular dipole moment $\boldsymbol{\mu}$ with respect to atomic displacements to obtain the atomic polar tensor (APT) \cite{Porezag1996,Cioslowski1989}
\begin{equation}
 A^{*}_{\kappa,\alpha\beta}
 =\frac{\partial\mu_{\alpha}}{\partial u_{\kappa\beta}},
 \label{eq:molecular_apt}
\end{equation}
in units of the elementary charge. Its rotationally invariant generalized APT charge is $q^{\mathrm{GAPT}}_{\kappa}=\mathrm{Tr}(\boldsymbol{A}^{*}_{\kappa})/3$. For a neutral molecule, a rigid translation cannot change the dipole moment, and therefore $\sum_{\kappa}\boldsymbol{A}^{*}_{\kappa}=\boldsymbol{0}$.

The molecular APT is the isolated limit of the BEC. For a molecule with negligible interactions between its periodic images, the electronic charge centers and the ionic coordinates give $\Omega\Delta\boldsymbol P=\Delta\boldsymbol\mu$ on a continuous polarization branch \cite{KingSmith1993}. Differentiating this relation at fixed cell gives $A^{*}_{\kappa,\alpha\beta}=eZ^{*}_{\kappa,\alpha\beta}$. Thus both tensors have the same numerical components when reported in units of $e$.

\subsection{Phonon modes and dielectric response}
\label{sec:phonon_dielectric}

Zone-center optical modes determine the phonon dielectric response. ZStar diagonalizes the mass-weighted matrix from $\Phi^\Gamma$ or reads the modes from a compatible independent phonon calculation.

For a normal mode $\lambda$ with angular frequency $\omega_{\lambda}$ and normalized eigendisplacements $e^{\lambda}_{\kappa\beta}$, the mode effective charge along $\alpha$ is \cite{Gonze1997,Veithen2005}
\begin{equation}
	Z_{\lambda,\alpha}
	=
	e
	\sum_{\kappa\beta}
	\frac{
		Z^{*}_{\kappa,\alpha\beta}
		e^{\lambda}_{\kappa\beta}
	}{
		\sqrt{M_{\kappa}}
	},
	\label{eq:mode_charge}
\end{equation}
Here $M_{\kappa}$ is the atomic mass. The factor $M_\kappa^{-1/2}$ converts the mass-weighted mode eigenvector to an atomic displacement per unit normal coordinate. Only modes with nonzero mode charge contribute to the infrared dielectric response. The corresponding selection rules for the 32 crystallographic point groups are given in \ref{app:point_group_phonons}.

In SI units, the static lattice contribution to the relative dielectric tensor is \cite{Gonze1997,Wu2005}
\begin{equation}
	\epsilon^{\mathrm{ph}}_{\alpha\beta}(0)
	=
	\frac{1}{\varepsilon_0\Omega}
	\sum_{\lambda \in \mathrm{IR}}
	\frac{
		Z_{\lambda,\alpha}
		Z_{\lambda,\beta}
	}{
		\omega_{\lambda}^{2}
	},
	\label{eq:lattice_dielectric_static}
\end{equation}
Here $\varepsilon_0$ is the vacuum permittivity and the sum runs over infrared-active zone-center modes. We use SI units. The total static dielectric tensor is
\begin{equation}
	\epsilon_{\alpha\beta}(0)
	=
	\epsilon^{\infty}_{\alpha\beta}
	+
	\epsilon^{\mathrm{ph}}_{\alpha\beta}(0),
	\label{eq:static_total_dielectric}
\end{equation}
where $\epsilon^\infty_{\alpha\beta}$ is the electronic dielectric tensor calculated with the optical module in PYATB \cite{Jin2023}.

At finite frequency, the same mode charges enter the damped oscillator sum \cite{Gonze1997,Baroni2001,Veithen2005},
\begin{equation}
	\epsilon^{\mathrm{ph}}_{\alpha\beta}(\omega)
	=
	\frac{1}{\varepsilon_0\Omega}
	\sum_{\lambda \in \mathrm{IR}}
	\frac{
		Z_{\lambda,\alpha}
		Z_{\lambda,\beta}
	}{
		\omega_{\lambda}^{2}
		-
		\omega^{2}
		-
		i \gamma_{\lambda}\omega
	},
	\label{eq:freq_lattice_dielectric}
\end{equation}
where $\gamma_{\lambda}$ is a phenomenological damping factor for mode $\lambda$. The total dielectric function is
\begin{equation}
	\epsilon_{\alpha\beta}(\omega)
	=
	\epsilon^{\infty}_{\alpha\beta}
	+
	\epsilon^{\mathrm{ph}}_{\alpha\beta}(\omega).
	\label{eq:total_dielectric_freq}
\end{equation}

\subsection{Piezoelectric response}
\label{sec:piezoelectric_theory}

Piezoelectricity couples homogeneous strain to polarization. We consider an insulating bulk crystal at zero macroscopic electric field and vanishing reference stress, using engineering Voigt strain
\begin{equation}
 \boldsymbol{\eta}
 =(\varepsilon_{xx},\varepsilon_{yy},\varepsilon_{zz},
 2\varepsilon_{yz},2\varepsilon_{xz},2\varepsilon_{xy})^{\mathsf T}.
 \label{eq:piezo_voigt}
\end{equation}
The piezoelectric stress coefficient $e_{\alpha\mu}$ has polarization index $\alpha$ and engineering strain index $\mu=1,\ldots,6$ \cite{Wu2005,Veithen2005}.

The proper piezoelectric response is the strain derivative of the polarization expressed in the reference cell \cite{Vanderbilt2000}. For deformation gradient $\mathbf F_{\mathrm{def}}$ and $J=\det\mathbf F_{\mathrm{def}}$, this polarization and its derivative are
\begin{equation}
 \boldsymbol{\mathcal P}=J\mathbf F_{\mathrm{def}}^{-1}\mathbf P,
 \qquad
 e_{\alpha\mu}
 =\left.\frac{\partial\mathcal P_\alpha}{\partial\eta_\mu}
 \right|_{\boldsymbol{\eta}=0,\,\mathbf E=0}.
 \label{eq:piezo_proper_definition}
\end{equation}
The corresponding infinitesimal correction to $e^{\mathrm{raw}}_{\alpha\mu}=\partial P_\alpha/\partial\eta_\mu$ is
\begin{align}
 e_{\alpha,(jj)}
 &=e^{\mathrm{raw}}_{\alpha,(jj)}+P_\alpha
   -\delta_{\alpha j}P_j,
 \label{eq:piezo_proper_normal}\\
 e_{\alpha,(jk)}
 &=e^{\mathrm{raw}}_{\alpha,(jk)}
   -\tfrac12(\delta_{\alpha j}P_k+\delta_{\alpha k}P_j),
 \qquad j\ne k.
 \label{eq:piezo_proper_shear}
\end{align}
Here $(jj)$ denotes a normal Voigt column, $(jk)$ denotes its engineering-shear counterpart, and the reference polarization is taken on the same continuous branch as the strain derivatives. The factor $1/2$ in the geometric correction follows from the definition of engineering shear in Eq.~(\ref{eq:piezo_voigt}). The raw coefficient is already a derivative with respect to the engineering shear strain. Phase-wrapped matching is performed on reduced polarization phases because the polarization quantum changes with strain. The combined derivative and correction are invariant under a consistent change of polarization branch \cite{Vanderbilt2000}.

Clamped-ion coefficients keep fractional atomic coordinates fixed under homogeneous deformation. Relaxed-ion coefficients additionally include the non-affine internal displacements $u^{\mathrm{int}}_{\kappa\beta}$. Their strain derivatives and contribution to the piezoelectric tensor are \cite{Wu2005}
\begin{equation}
 \Lambda_{\kappa\beta\mu}
 =\frac{\partial u^{\mathrm{int}}_{\kappa\beta}}{\partial\eta_\mu},
 \qquad
 e^{\mathrm{rel}}_{\alpha\mu}
 =e^{\mathrm{cl}}_{\alpha\mu}
 +\frac{e}{\Omega}
   \sum_{\kappa\beta}Z^*_{\kappa,\alpha\beta}
   \Lambda_{\kappa\beta\mu}.
 \label{eq:piezo_internal_relaxation}
\end{equation}
Measure the internal displacement from the atomic position obtained by applying the homogeneous strain. We remove the common translation from the internal displacements. Its contribution vanishes when the BEC acoustic sum rule holds.

The same zone-center force constants introduced in Section~\ref{sec:unified_response} also determine internal relaxation. Defining the clamped-ion force--strain coupling $\Xi_{\kappa\beta\mu}=\partial F_{\kappa\beta}/\partial\eta_\mu$, the harmonic force balance is $\mathbf F=-\boldsymbol{\Phi}^{\Gamma}\mathbf u^{\mathrm{int}}+\boldsymbol{\Xi}\boldsymbol{\eta}$. Thus
\begin{equation}
 \boldsymbol{\Lambda}
 =(\boldsymbol{\Phi}^{\Gamma})^+\boldsymbol{\Xi},
 \qquad
 \mathbf C^{E,\mathrm{rel}}
 =\mathbf C^{E,\mathrm{cl}}
 -\frac{1}{\Omega}\boldsymbol{\Xi}^{\mathsf T}
  (\boldsymbol{\Phi}^{\Gamma})^+\boldsymbol{\Xi}.
 \label{eq:piezo_harmonic_relaxation}
\end{equation}
The pseudoinverse is evaluated in the optical subspace, excluding rigid translations, with consistently converted force and length units. The polarization and stress of the strained cells give the clamped-ion piezoelectric and elastic coefficients \cite{Wu2005}.

Finally, the strain coefficients at free mechanical stress follow from the fixed-electric-field elastic compliance:
\begin{equation}
 \mathbf d=\mathbf e^{\mathrm{rel}}\mathbf S^{E,\mathrm{rel}},
 \qquad
 \mathbf S^{E,\mathrm{rel}}
 =(\mathbf C^{E,\mathrm{rel}})^{-1}.
 \label{eq:piezo_d_conversion}
\end{equation}
The corresponding dimensional conversion is
\begin{equation}
 \frac{\mathrm C}{\mathrm m^2}\frac{\mathrm m^2}{\mathrm N}
 =\frac{\mathrm C}{\mathrm N}=\frac{\mathrm m}{\mathrm V},
 \qquad
 \mathbf d\,[\mathrm{pm/V}]
 =10^3\,\mathbf e\,[\mathrm{C/m^2}]
  (\mathbf C^E\,[\mathrm{GPa}])^{-1}.
 \label{eq:piezo_unit_conversion}
\end{equation}
For insulating bulk crystals, ZStar uses engineering shear and elastic compliance at the same ionic and electrical boundary conditions as the proper piezoelectric response. The output strain coefficients can be expressed equivalently in pC/N or pm/V.

\subsection{Phonon dispersion, LO--TO splitting, and the non-analytic correction}
\label{sec:phonon_nac_theory}

The force constants from finite displacements in the input cell give
the analytic dynamical matrix at the zone center. A phonon dispersion is
obtained by Fourier transforming force constants from a real-space supercell.
For a polar bulk crystal, an optical displacement produces a macroscopic
polarization through the BEC tensors. The resulting long-range dipole--dipole
interaction is non-analytic as $\mathbf q\rightarrow\mathbf 0$; its directional
limit shifts a longitudinal optical (LO) mode relative to the transverse
optical (TO) limit. This is the microscopic origin of the LO--TO splitting and
requires a non-analytic term in addition to the short-range force constants
\cite{Gonze1997,Baroni2001,Zhong1994}:
\begin{equation}
 D_{\kappa\alpha,\kappa'\beta}(\mathbf q)
 =D^{\mathrm A}_{\kappa\alpha,\kappa'\beta}(\mathbf q)
 +D^{\mathrm{NA}}_{\kappa\alpha,\kappa'\beta}(\hat{\mathbf q}).
 \label{eq:nac_dynamical_matrix}
\end{equation}
With the polarization-first BEC convention used throughout this work, define
\(\mathcal Z_{\kappa\alpha}(\hat{\mathbf q})=
\sum_{\gamma}\hat q_{\gamma}Z^{*}_{\kappa,\gamma\alpha}\). The non-analytic
term in the limit $\mathbf q\to\mathbf 0$ along fixed $\hat{\mathbf q}$ is
\begin{equation}
 D^{\mathrm{NA}}_{\kappa\alpha,\kappa'\beta}(\hat{\mathbf q})
 =\frac{e^2}{\varepsilon_0\Omega}
 \frac{\mathcal Z_{\kappa\alpha}(\hat{\mathbf q})
       \mathcal Z_{\kappa'\beta}(\hat{\mathbf q})}
 {\sqrt{M_\kappa M_{\kappa'}}\,\sum_{\gamma\delta}\hat q_{\gamma}
  \epsilon^{\infty}_{\gamma\delta}\hat q_{\delta}}.
 \label{eq:nac_term}
\end{equation}
Here \(\Omega\) is the primitive-cell volume, \(M_\kappa\) is the atomic
mass, and \(\epsilon^{\infty}\) is the electronic dielectric tensor. The
directional denominator in Eq.~(\ref{eq:nac_term}) expresses the screening of
the macroscopic field, while the two BEC factors describe how the displaced
atoms couple to that field. An uncorrected zone-center mode with zero mode-effective charge projection along
the approach direction has zero correction.
In an anisotropic crystal, the zone-center LO frequency
can vary with the direction of approach.

For bulk crystals, ZStar combines supercell force constants with the directional zone-center correction using the BECs and electronic dielectric tensor in a Phonopy-compatible \texttt{BORN} file.
Slabs, nanowires, and molecules require different Coulomb boundary conditions. The optional
finite-wavevector workflow uses Phonopy to generate the supercell
displacements, collect force constants, and evaluate the standardized
high-symmetry path and phonon density of states \cite{Togo2015,Togo2023}.

\subsection{Infrared and Raman spectra}
\label{sec:raman_theory}
The mode charges in Eq.~(\ref{eq:mode_charge}) set the infrared oscillator strengths. For bulk crystals, Eqs.~(\ref{eq:freq_lattice_dielectric}) and (\ref{eq:total_dielectric_freq}) give the relative permittivity. For isolated molecules, contracting the BECs with the normal modes gives the dipole derivatives directly. The same modes project the polarizability derivatives used for molecular Raman intensities.

For a slab, division by a supercell volume introduces an explicit vacuum-volume factor. ZStar removes this factor and reports the source-field-normalized sheet response
\begin{equation}
 \boldsymbol{\alpha}^{2\mathrm{D}}(\omega)/\varepsilon_0
 =
 \frac{1}{\varepsilon_0 A}
 \sum_{\lambda}
 \frac{
 \boldsymbol{Z}_{\lambda}\otimes\boldsymbol{Z}_{\lambda}
 }{
 \omega_{\lambda}^{2}-\omega^{2}-i\gamma_{\lambda}\omega
 }
 +
 L_z\left(\boldsymbol{\epsilon}^{\infty}-\boldsymbol{I}\right),
 \label{eq:2d_ir_sheet}
\end{equation}
Here $\alpha^{2\mathrm D}$ is the polarizability per area, and $\alpha^{2\mathrm D}/\varepsilon_0$ has units of length and equals $4\pi$ times the Gaussian sheet polarizability. PYATB's optical and direct-static kernels give the independent-particle response to the applied source field. For a chosen thickness $t$, the equivalent relative permittivity in this convention is $\boldsymbol{I}+\boldsymbol{\alpha}^{2\mathrm D}/(\varepsilon_0t)$.

For a diagonal slab tensor obtained from a screened macroscopic permittivity $\epsilon^{\mathrm{SC}}$, the conversion follows from continuity of the tangential electric field and normal displacement field \cite{Laturia2018}
\begin{equation}
 \epsilon^{\mathrm{layer}}_{\parallel}
 =1+\frac{L_z}{t}(\epsilon^{\mathrm{SC}}_{\parallel}-1),
 \qquad
 \epsilon^{\mathrm{layer}}_{\perp}
 =\left[1+\frac{L_z}{t}\left(\frac{1}{\epsilon^{\mathrm{SC}}_{\perp}}-1\right)\right]^{-1}.
 \label{eq:slab_boundary_conversion}
\end{equation}
For a screened macroscopic supercell permittivity, ZStar also applies these boundary conditions to the full tensor. The following examples report the source-field-normalized sheet response.

For a wire of period $L_z$, the analogous line normalization is
\begin{equation}
 \boldsymbol{\alpha}^{1\mathrm D}(\omega)/\varepsilon_0
 =
 \frac{1}{\varepsilon_0 L_z}
 \sum_{\lambda}
 \frac{\boldsymbol{Z}_{\lambda}\otimes\boldsymbol{Z}_{\lambda}}
 {\omega_{\lambda}^{2}-\omega^{2}-i\gamma_{\lambda}\omega}
 +A_{\perp}\left(\boldsymbol{\epsilon}^{\infty}-\boldsymbol{I}\right),
 \label{eq:1d_ir_line}
\end{equation}
which has units of area. As for sheets, the conversion retains the source field convention. The width of the vacuum gap controls the interaction between periodic images.

In the ABACUS workflow, ZStar evaluates $\epsilon^\infty$ for the same Phonopy-displaced structures used to obtain the BECs. It subtracts the reference $\epsilon^\infty$, fits the displacement derivatives, and projects them onto Phonopy eigenvectors. The resulting bulk Raman tensor is the derivative along the mass-weighted mode coordinate $q_\lambda$ \cite{Umari2001Raman,Veithen2005,BastoneroMarzari2024},
\begin{equation}
 R_{\lambda,\alpha\gamma}=\frac{\partial\epsilon^\infty_{\alpha\gamma}}{\partial q_\lambda},
 \qquad u_{\kappa\beta} = q_\lambda\, e^\lambda_{\kappa\beta}/\sqrt{M_\kappa}.
 \label{eq:raman_tensor}
\end{equation}
The eigenvectors have unit norm. With $M_\kappa$ in atomic mass units, ZStar stores $q_\lambda$ in \AA$\sqrt{\mathrm{amu}}$. The bulk tensor uses relative permittivity. To express it as the Raman susceptibility in Eq.~(9) of Ref.~\cite{Umari2001Raman}, multiply it by $\sqrt{\Omega}/(4\pi)$, since their Gaussian susceptibility is $(\epsilon^\infty-I)/(4\pi)$. Table~\ref{tab:raman_units} gives the quantities differentiated for finite systems. The non-resonant powder activity follows the Placzek invariant \cite{Porezag1996},
\begin{equation}
 A_{\lambda}=45\bar{\alpha}_{\lambda}^{2}+7\gamma_{\lambda,\mathrm{aniso}}^{2},
 \qquad
 \bar{\alpha}_{\lambda}=\frac{1}{3}\mathrm{Tr}\,\boldsymbol{R}_{\lambda},
 \label{eq:placzek_invariant}
\end{equation}
where $\gamma_{\lambda,\mathrm{aniso}}^{2}$ is the standard anisotropic invariant of the symmetrized tensor and is distinct from the damping factor in Eq.~(\ref{eq:freq_lattice_dielectric}). The Stokes intensity includes the Bose factor and laser-frequency prefactor \cite{Umari2001Raman,Veithen2005},
\begin{equation}
 I_{\lambda}\propto
 \frac{(\tilde\nu_L-\tilde\nu_{\lambda})^{4}}
 {\tilde\nu_{\lambda}}
 \left[n(\tilde\nu_{\lambda},T)+1\right]A_{\lambda}.
\end{equation}
We use the symmetry-related displacements to fit $G_{\kappa,\alpha\gamma\beta}$, the derivative of the relative permittivity with respect to the displacement of atom $\kappa$. For observation $s$,
\begin{equation}
 \Delta\epsilon^{\infty,(s)}_{\alpha\gamma}
 =\sum_{\beta}G_{\kappa,\alpha\gamma\beta}u^{(s)}_{\kappa\beta}
 +\mathcal O(\lVert\boldsymbol u^{(s)}\rVert^2),\qquad
 G_{\kappa,\alpha\gamma\beta}
 =\frac{\partial\epsilon^{\infty}_{\alpha\gamma}}{\partial u_{\kappa\beta}}.
 \label{eq:unified_raman_observation}
\end{equation}
The factor $c_d$ in Eq.~(\ref{eq:unified_raman_projection}) converts $G$ to the atomic Raman response in the units of Table~\ref{tab:raman_units}. Under a symmetry operation with Cartesian rotation $\boldsymbol S$ and atom permutation $p$, the derivative transforms as
\begin{equation}
 G_{p(\kappa),\alpha\gamma\beta}
 =\sum_{a b c}S_{\alpha a}S_{\gamma b}S_{\beta c}
 G_{\kappa,a b c}.
 \label{eq:unified_raman_symmetry}
\end{equation}
Contraction with the mass-weighted eigenvectors then gives
\begin{equation}
 R_{\lambda,\alpha\gamma}
 =c_d\sum_{\kappa\beta}G_{\kappa,\alpha\gamma\beta}
 \frac{e^{\lambda}_{\kappa\beta}}{\sqrt{M_{\kappa}}},\qquad
 c_d=\begin{cases}
 1,&d=3,\\
 L_z,&d=2,\\
 A_\perp/(4\pi),&d=1,\\
 \Omega/(4\pi),&d=0.
 \end{cases}
 \label{eq:unified_raman_projection}
\end{equation}
The displaced-cell SCFs supply the forces and matrices for polarization and Raman calculations. PYATB uses the Hamiltonian, overlap, and position matrices to calculate the independent-particle electronic dielectric tensor through its Kubo--Greenwood response \cite{Jin2023}. The field derivative of forces is an alternative expression for the BEC \cite{Wu2005}.

The zero-field derivative in Eq.~(\ref{eq:unified_raman_observation}) gives the transverse optical response. In a polar bulk crystal, the macroscopic field of a longitudinal vibration changes both the eigenvectors and the dielectric derivative \cite{Veithen2005}. Equation~(\ref{eq:nac_term}) gives the dynamical correction. The additional optical term is $(\partial\epsilon^\infty_{\alpha\gamma}/\partial \mathcal E_c)\mathcal E_{\lambda,c}$, with the field per mode amplitude determined by its effective charge and directional dielectric screening. The \texttt{zstar dielectric electrooptic} command computes the ionic term of the fixed-strain Pockels tensor from the same zone-center mode charges and Raman tensors. The command takes the static electronic $\chi^{(2)}$ to calculate the clamped-ion Pockels term and the macroscopic-field correction to the Raman tensor. The electronic and ionic terms sum to the fixed-strain tensor. For free stress, the strain induced by the field adds $\sum_{\nu=1}^{6} p_{ij\nu}d_{\nu k}$ to the fixed-strain coefficient $r_{ijk}$. Here $p_{ij\nu}=\partial(\epsilon^\infty)^{-1}_{ij}/\partial\eta_\nu$ is evaluated after internal relaxation, and $d_{\nu k}$ gives the engineering strain induced by the field \cite{Veithen2005}.

For a field applied along a nonperiodic direction $b$, the self-consistent response gives an alternative Raman derivative \cite{Porezag1996,Veithen2005}:
\begin{equation}
 R_{\lambda,ab}^{(d),\mathrm{FF}}
 =\left.\frac{c_d}{\varepsilon_0}
 \frac{\partial^2 P_a^{\mathrm{SCF}}(q_\lambda,\mathcal E_b)}
 {\partial q_\lambda\,\partial\mathcal E_b}
 \right|_{q_\lambda=\mathcal E_b=0},
 \label{eq:raman_open_finite_field}
\end{equation}
Here $P_a$ is the supercell polarization from real-space dipoles for open $a$ and Berry phases for periodic $a$. Each applied field at a displaced geometry uses its own self-consistent electronic density. ZStar uses Eq.~(\ref{eq:raman_open_finite_field}) for the two transverse directions of the chain. The axial $R_{\lambda,zz}$ comes from the zero-field derivative.

\begin{table}[htbp]
\centering\small
\caption{Raman normalization in ZStar. $\mathcal X$ is the differentiated electronic response, and $\boldsymbol{R}_\lambda=\partial\mathcal X/\partial q_\lambda$. Tensor units show the normalization of $\mathcal X$.}
\label{tab:raman_units}
\begin{tabular}{lllll}
\toprule
Dimensionality & System & $\mathcal X$ & Unit of $\mathcal X$ & Unit of $\boldsymbol{R}_\lambda$ \\
\midrule
3D & Bulk & $\epsilon^\infty$ & $1$ & $1/(\text{\AA}\sqrt{\mathrm{amu}})$ \\
2D & Two-dimensional material & $L_z(\epsilon^\infty-I)$ & \AA & $1/\sqrt{\mathrm{amu}}$ \\
1D & Nanowire & $A_\perp(\epsilon^\infty-I)/(4\pi)$ & \AA$^2$ & \AA$/\sqrt{\mathrm{amu}}$ \\
0D & Molecule & $\Omega(\epsilon^\infty-I)/(4\pi)$ & \AA$^3$ & \AA$^2/\sqrt{\mathrm{amu}}$ \\
\bottomrule
\end{tabular}
\end{table}

\vspace*{-5mm}

\section{Software}
\label{sec:software}

\clubpenalty=10000
\widowpenalty=10000
\displaywidowpenalty=10000

\subsection{Implementation and commands}
A calculation starts from a relaxed primitive cell. The ABACUS files \texttt{INPUT}, \texttt{STRU}, and \texttt{KPT} specify the calculation settings, atomic structure, and $k$-point mesh, respectively. The command \texttt{zstar config check} verifies the executable path and basis settings. For the BEC calculation, the command sequence is
\begin{lstlisting}[language=bash]
zstar bec pre --stru STRU -i INPUT
zstar bec run
zstar bec stat
zstar bec post
\end{lstlisting}
The first command prepares the reference cell and finite displacements. The second runs ABACUS and PYATB. \texttt{stat} reports progress, and \texttt{post} writes the BECs, Phonopy \texttt{BORN} file, and zone-center modes. For a cluster, use \texttt{zstar bec job --system slurm} to generate the launch script. The BaTiO$_3$ example in Section~\ref{sec:run_example} illustrates input preparation.

The \texttt{zstar} command uses the actions \texttt{pre}, \texttt{job}, \texttt{run}, \texttt{stat}, and \texttt{post} for its main calculation families (Table~\ref{tab:zstar_commands}). Preparation writes the calculator, dimensionality, structure location, and options to a family configuration such as \path{.zstar/bec.json}. The files in \path{.zstar/stages/} record stage status and errors, and \path{.zstar/workflow.jsonl} records events in execution order. The optional \texttt{job} action writes the launch command to a shell, Slurm, or Torque driver. Raw calculator outputs remain in the calculation directories.

\begin{table}[h]
 \centering

\caption{ZStar commands by calculation family.}
 \label{tab:zstar_commands}
 \footnotesize
 \begin{tabularx}{\textwidth}{>{\raggedright\arraybackslash}p{0.46\textwidth}X}
  \toprule
  \textbf{Command family} & \textbf{Purpose} \\
  \midrule
  \mbox{\texttt{zstar bec pre/job/run/stat/post}} & Calculate polarization, BECs, and zone-center force constants. \\
  \mbox{\texttt{zstar phonon pre/job/run/stat/post/irrep/spectrum}} & Calculate phonon dispersions and DOS, and classify zone-center modes. \\
  \mbox{\texttt{zstar spectra pre/job/run/stat/post}} & Calculate infrared and Raman spectra. \\
  \mbox{\texttt{zstar piezo pre/job/run/stat/post}} & Calculate bulk piezoelectric and elastic tensors. \\
  \mbox{\texttt{zstar dielectric static/freq/optics}} & Calculate static and frequency-dependent dielectric responses. \\
  \mbox{\texttt{zstar backend list}} & List capabilities for ABACUS, VASP, CP2K, Quantum ESPRESSO, etc. \\
  \mbox{\texttt{zstar config init/show/set/check}} & Configure executables, parallel settings, and ABACUS basis files. \\
  \mbox{\texttt{zstar response}} & Import response records/prepare predicted Hamiltonians for PYATB. \\
  \mbox{\texttt{zstar density}} & Prepare charge-density exports and record their conventions. \\
  \mbox{\texttt{zstar stru convert/wyckoff}} & Convert structures and list symmetry representatives. \\
  \mbox{\texttt{zstar data db collect}} & Collect BEC results in CSV and JSONL format. \\
  \mbox{\texttt{zstar skill install/path/preflight}} & Install, locate, or run the preflight check for the agent skill. \\
  \mbox{\texttt{zstar pot}} & Plot electrostatic potentials, vacuum steps, and mirror asymmetry. \\
  \bottomrule
 \end{tabularx}
\end{table}

\FloatBarrier

\subsection{Dependencies and installation}

The package declares support for Python 3.9 and later and depends on NumPy \cite{Harris2020NumPy}, SciPy \cite{Virtanen2020SciPy}, PyYAML, Matplotlib \cite{Hunter2007Matplotlib}, spglib \cite{Togo2024Spglib}, and Phonopy \cite{Togo2015,Togo2023}. ZStar reads ABACUS \texttt{STRU} files directly. The optional pymatgen dependency \cite{Ong2013Pymatgen} supports VASP \texttt{vasprun.xml} and \texttt{CHGCAR}/\texttt{POTCAR} inputs. ABACUS \cite{Li2016,ABACUS2025} supplies Hamiltonian matrices, charge density, total energy, and forces. PYATB \cite{Jin2023} supplies Berry-phase polarization, band data, and electronic dielectric response, while Phonopy generates displacements and calculates phonons. ZStar also provides calculator-specific adapters for VASP \cite{Kresse1996VASP}, CP2K \cite{Kuhne2020CP2K}, and Quantum ESPRESSO \cite{Giannozzi2009,Giannozzi2017}. The command \texttt{zstar backend list} lists the response capabilities declared by each adapter.

\Needspace{8\baselineskip}
Install ZStar from PyPI with:
\begin{lstlisting}[language=bash]
pip install zstar
zstar config check
zstar backend list --check
\end{lstlisting}
External electronic-structure programs are installed separately. ZStar and PYATB should be available in the same Python environment for the unified ABACUS route. Executable paths and MPI/OpenMP settings can be configured for the user or the current project, with command-specific overrides. ABACUS pseudopotential and orbital directories may be supplied through \texttt{--pp /path/to/PSEUDO --orb /path/to/ORBITAL}. During preparation, ZStar copies the basis files referenced by \texttt{STRU}. If several files match, it reports the ambiguity.

\Needspace{8\baselineskip}
\subsection{Unified displacement-response workflow}
\label{sec:unified_workflow}

The bulk \mbox{ABACUS+PYATB} calculation uses the finite-displacement method in Section~\ref{sec:unified_response}. Run it with
\begin{lstlisting}[language=bash]
zstar bec pre --stru STRU
zstar bec run
zstar bec post
zstar dielectric static
\end{lstlisting}
Preparation can generate an initial SCF input. Use \texttt{-i INPUT} to supply the user's SCF settings. Physical dimensionality is selected at the same stage using \texttt{--dim 2} for a two-dimensional material, \texttt{--dim 1} for a nanowire, or \texttt{--dim 0} for a molecule. The corresponding dimensionality must also be supplied to dielectric post-processing.
\paragraph{Preparation}
The symmetry-adapted displacement directions in the input cell are generated by Phonopy. A negative displacement is added when the site symmetry does not supply the opposite direction. The default displacement is $0.02$~bohr ($\sim0.01$~\AA). The reference is in \texttt{0.no-move}, and the displaced structures are in \texttt{disp-001}, \texttt{disp-002}, and subsequent directories. ZStar reads the Cartesian displacements from those structures and lists the stages in \texttt{shared\_response.json}.

\paragraph{Workflow execution}
After the reference SCF calculation, ZStar checks the gap and band occupations. The displaced calculations start from the reference charge density, and polarization uses the reference number of occupied bands. The same runs supply the forces. PYATB evaluates the reference electronic dielectric tensor once. Use \texttt{zstar bec stat} to find unfinished calculations and resume with \texttt{zstar bec run}.

\paragraph{Response collection}
ZStar matches the polarization branches, combines the periodic and open-direction responses, and subtracts the reference forces. It checks the rank of each site fit and expands the fitted tensors by symmetry. ZStar saves the BECs and force constants before and after applying sum rules. The corrected tensors and zone-center modes feed the dielectric and infrared calculations.

\paragraph{Independent and subsequent calculations}
The Unified workflow calculates zone-center force constants in the input cell. For finite-wavevector phonons, run \texttt{zstar phonon} with a supercell.

In a prepared Unified directory, \texttt{zstar spectra pre/run/post} contracts BECs with the phonon eigenvectors to obtain IR intensities. For nonresonant Raman spectra, it reuses the Hamiltonian, overlap, and position matrices from the BEC calculations to evaluate the dielectric tensor, then fits and projects its displacement derivatives onto the normal modes. Use \texttt{--response PATH} to select a different prepared displacement set or \texttt{--method mode} to perform independent normal-mode finite differences. Table~\ref{tab:raman_units} gives the Raman units and normalizations. For cluster runs, a Slurm script can be generated with \texttt{zstar bec job --system slurm}. Scheduler settings are described in the manual.

\Needspace{8\baselineskip}
\subsection{Compatibility and output}
\label{sec:model_response_usage}
\label{sec:response_outputs}

ZStar reads electronic and response data from ABACUS+PYATB, VASP, CP2K, and Quantum ESPRESSO. Examples for the calculator adapters are in \path{examples/backend_examples/}. It also accepts Hamiltonians predicted by HamGNN \cite{Zhong2023HamGNN}, DeepH \cite{Li2022DeepH}, NextHAM \cite{Yin2026NextHAM}, and Hot-Ham \cite{Liang2026HotHam}. ZStar converts real Hamiltonians in the ABACUS basis from HamGNN and DeepH into PYATB format. NextHAM and Hot-Ham export ABACUS sparse matrices that PYATB reads directly. With the overlap and position matrices in the same basis and atom order, PYATB can evaluate polarization for the reference and displaced structures \cite{Jin2023}. Model inputs and results are in \path{examples/predicted_hamiltonians/}. Section~\ref{sec:field_aware_ml} illustrates how the calculated responses are used for training.

The common output \texttt{response.json} gives the tensors, units, and axis conventions. For a displacement calculation, \texttt{BEC.raw.dat} and \texttt{BEC.dat} give the raw and charge-neutral BECs. The joint solution records the polarization and force observations in \texttt{response\_fit.json}. \texttt{FORCE\_SETS} gives the forces and displacements, and \texttt{qpoints.yaml} gives the phonon frequencies and eigenvectors. In the Phonopy \texttt{BORN} file, the first tensor index denotes the polarization direction. ZStar uses the force-atom-first order in Eq.~(\ref{eq:unified_force_observations}). Bulk strain calculations report the proper piezoelectric tensor $\mathbf e$, the fixed-field elastic tensor $\mathbf C^E$, and $\mathbf d=\mathbf e(\mathbf C^E)^{-1}$ \cite{Wu2005,Kresse1996VASP,deJong2015Piezo}. The complex electronic dielectric tensor also gives absorption, reflectivity, and the loss function.

\subsection{Command-line and agent skill}

ZStar includes the \texttt{zstar-workflows} skill for researchers who use coding agents. A researcher can describe the target calculation in plain language, such as obtaining BECs and zone-center phonons for a relaxed bulk structure. The skill identifies the ZStar command sequence and the required inputs for that workflow. It guides the agent through preparing the calculation, checking progress, and locating the resulting tensors and spectra.

\begin{lstlisting}[language=bash]
zstar skill install
zstar skill preflight --lane bec --dim bulk
\end{lstlisting}
The preflight command checks the input files and calculator executables before a run. Users can also invoke the same calculation commands directly.

\section{Examples and Benchmarks}
\label{sec:examples}

We first test BECs across four dimensionalities, using bulk BaTiO$_3$ and HfO$_2$, two-dimensional hBN and $\alpha$-In$_2$Se$_3$, the BN(9,0) nanotube and Sb$_2$S$_3$ chain, and the H$_2$O and CH$_4$ molecules. The following examples show how these responses enter dielectric tensors, phonon dispersions, and IR and Raman spectra. We compare bulk piezoelectric tensors with published values, use the BaTiO$_3$ dataset to train a field-aware model, and measure the efficiency of the unified workflow. The atomic structures were visualized with VESTA \cite{Momma2011VESTA}.

\subsection{Preparing and running the unified workflow}
\label{sec:run_example}
The ABACUS response calculations use numerical atomic orbital basis and atomic displacements of 0.02~Bohr ($\sim$0.01~\AA). The structures, $k$-point meshes, basis sets, and solver inputs are provided under \path{examples/}.

Here, we use the BEC calculation for cubic $Pm\bar{3}m$ BaTiO$_3$ to illustrate the setup. Its \texttt{INPUT} file selects PBEsol functional. To prepare the response calculation, run
\texttt{zstar bec pre --stru STRU -i INPUT}. The essential preparation tree is as below:
\begin{lstlisting}[language=bash]
./
|-- 0.no-move/
|-- disp-001/
|-- disp-002/
|-- disp-003/
|-- shared_response.json
|-- phonopy_disp.yaml
|-- INPUT
|-- KPT
+-- STRU
\end{lstlisting}

The three displaced structures correspond to Ba along $x$ (\texttt{disp-001}), Ti along $x$ (\texttt{disp-002}), and O along $x+y$ (\texttt{disp-003}), each with a displacement norm of 0.02~Bohr. The oxygen seed has both longitudinal and transverse components relative to its Ti--O bond; the site-symmetry orbit spans all three Cartesian directions. Each stage contains its own \texttt{INPUT}, \texttt{STRU},  \texttt{KPT}, and required basis files. The packaged runner supplies \texttt{--pp assets --orb assets} to resolve the included basis files automatically. Every displacement was initialized from the reference charge density copied into its ABACUS \texttt{OUT.<suffix>} directory.

The examples use functional of PBE \cite{Perdew1996PBE} or PBEsol \cite{Perdew2008}. For calculations that include a vdW correction, the input files specify whether the D3 correction uses zero damping \cite{Grimme2010D3} or Becke--Johnson damping \cite{Grimme2011BJ}. Table~\ref{tab:fresh_workflow_gaps} lists the functionals, symmetries, and reference gaps of the eight BEC examples. 

\begin{table*}[h]
	\centering
	\footnotesize
	\caption{Calculation settings and reference-state gaps for the BEC examples. Symmetry refers to the space group of the simulation cell for periodic systems and to the point group for molecules. The periodic values are PYATB gaps along the sampled band paths. The molecular values are occupied--unoccupied orbital gaps.}
	\label{tab:fresh_workflow_gaps}
	\setlength{\tabcolsep}{5pt}
	\begin{tabularx}{\textwidth}{>{\raggedright\arraybackslash}p{0.135\textwidth}*{8}{>{\centering\arraybackslash}X}}
		\toprule
		System & c-BaTiO$_3$ & t-HfO$_2$ & hBN & $\alpha$-In$_2$Se$_3$ & BN(9,0) & Sb$_2$S$_3$ & H$_2$O & CH$_4$ \\
		\midrule
		Symmetry & $Pm\bar{3}m$ & $P4_2/nmc$ & $P\bar{6}m2$ & $P3m1$ & $Pmc2_1$ & $P2_1/m$ & $C_{2v}$ & $T_d$ \\
		Dimension & 3D & 3D & 2D & 2D & 1D & 1D & 0D & 0D \\
		XC & PBEsol & PBEsol & PBE & PBEsol & PBE & PBE & PBE & PBE \\
		VdW correction& -- & -- & -- & D3(0) & -- & D3(BJ) & -- & -- \\
		Gap (eV) & 1.686 & 4.710 & 4.673 & 0.840 & 3.813 & 1.501 & 6.569 & 9.535 \\

		\bottomrule
	\end{tabularx}
\end{table*}

\subsection{Bulk BEC: \texorpdfstring{BaTiO$_3$ and HfO$_2$}{BaTiO3 and HfO2}}

Figure~\ref{fig:bulk_bec_structures}(a) shows cubic $Pm\bar{3}m$ BaTiO$_3$, whose symmetry makes the Ba and Ti diagonal responses isotropic and separates the oxygen response into Ti--O-bond-parallel and perpendicular components. Figure~\ref{fig:bulk_bec_structures}(b) shows tetragonal $P4_2/nmc$ HfO$_2$, where the unique $c$ axis and inequivalent local oxygen directions motivate the component-resolved comparison below. Both views use the geometries retained with the BEC calculations.

\begin{center}
\begin{minipage}{\textwidth}
 \centering
 \includegraphics[width=0.75\textwidth]{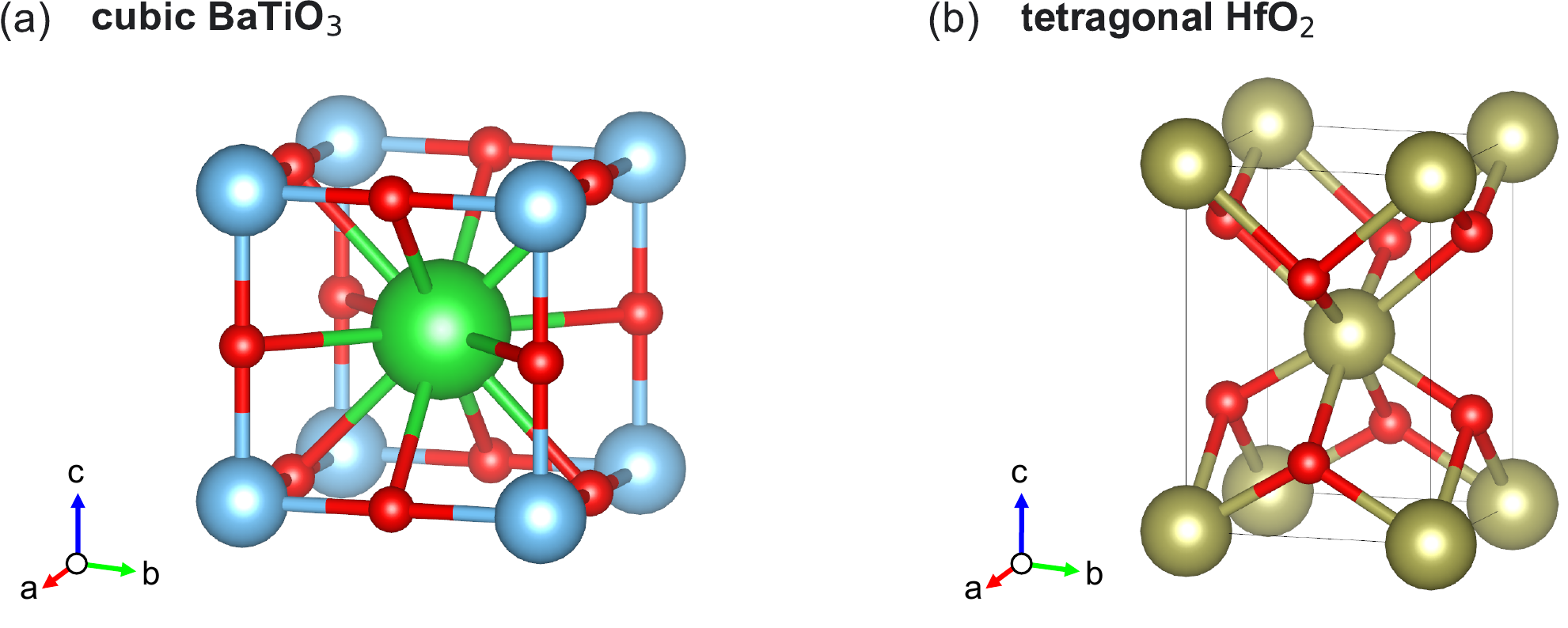}
 \captionsetup{hypcap=false}
 \captionof{figure}{Bulk benchmark structures. (a) Cubic $Pm\bar{3}m$ BaTiO$_3$. (b) Tetragonal $P4_2/nmc$ HfO$_2$.}
 \label{fig:bulk_bec_structures}
\end{minipage}
\end{center}

The unified cubic BaTiO$_3$ calculation yields charge-neutral tensors with $Z^*_{\mathrm{Ba}}=2.734$, $Z^*_{\mathrm{Ti}}=7.440$, $Z^*_{\mathrm{O},\parallel}=-5.861$, and $Z^*_{\mathrm{O},\perp}=-2.156$. These values reproduce the anomalous Ti--O response found in previous calculations for the cubic phase (Table~\ref{tab:bto_bec_literature}) \cite{Bilc2008,Masuki2022}. For tetragonal HfO$_2$, the corrected representative tensors are
\begin{equation}
 \boldsymbol{Z}^*_{\mathrm{Hf}}=
 \begin{pmatrix}
  5.394 & 0 & 0\\
  0 & 5.394 & 0\\
  0 & 0 & 4.828
 \end{pmatrix},
 \qquad
 \boldsymbol{Z}^*_{\mathrm O}=
 \begin{pmatrix}
  -2.115 & 0 & 0\\
  0 & -3.278 & 0\\
  0 & 0 & -2.414
 \end{pmatrix}.
\end{equation}
Table~\ref{tab:hfo2_bec_literature} compares these PBEsol tensors with our VASP results and Phonix \cite{Ohnishi2026Phonix}. The VASP results agree with Phonix to within $0.0075\,e$ for the Hf and O tensor components.

\begin{table*}[!htbp]
 \centering
 \footnotesize
 \caption{BECs of cubic BaTiO$_3$ in units of $e$. Cubic symmetry makes the Ba and Ti diagonal components isotropic. For oxygen, $\parallel$ and $\perp$ denote directions parallel and perpendicular to the Ti--O bond, respectively. All columns refer to the cubic phase.}
 \label{tab:bto_bec_literature}
 \begin{tabularx}{\textwidth}{>{\raggedright\arraybackslash}p{0.18\textwidth}>{\raggedright\arraybackslash}p{0.16\textwidth}>{\centering\arraybackslash}p{0.08\textwidth}*{4}{>{\centering\arraybackslash}X}}
  \toprule
  Source & Software & XC & Ba & Ti & O$_{\parallel}$ & O$_{\perp}$ \\
  \midrule
  ZStar & ABACUS+PYATB & PBEsol & 2.734 & 7.440 & $-5.861$ & $-2.156$ \\
  Masuki \textit{et al.}\ \cite{Masuki2022} & VASP & PBEsol & 2.725 & 7.068 & $-5.576$ & $-2.109$ \\
  Bilc \textit{et al.}\ \cite{Bilc2008} & CRYSTAL & PBE & 2.730 & 7.170 & $-5.730$ & $-2.080$ \\

  \bottomrule
 \end{tabularx}
\end{table*}

\begin{table*}[htbp]
 \centering
 \small
 \caption{BECs of tetragonal HfO$_2$ in units of $e$. The oxygen components are given in a common symmetry-matched local frame.}
 \label{tab:hfo2_bec_literature}
 \begin{tabularx}{\textwidth}{>{\raggedright\arraybackslash}p{0.19\textwidth}>{\raggedright\arraybackslash}p{0.15\textwidth}>{\centering\arraybackslash}p{0.08\textwidth}*{5}{>{\centering\arraybackslash}X}}
  \toprule
  Source & Software & XC & Hf($xx$) & Hf($zz$) & O($xx$) & O($yy$) & O($zz$) \\
  \midrule
  ZStar & ABACUS+PYATB & PBEsol & 5.394 & 4.828 & $-2.115$ & $-3.278$ & $-2.414$ \\
  ZStar & VASP (DFPT) & PBEsol & 5.513 & 4.866 & $-2.127$ & $-3.386$ & $-2.433$ \\
  Phonix \cite{Ohnishi2026Phonix} & VASP & PBEsol & 5.509 & 4.873 & $-2.131$ & $-3.378$ & $-2.437$ \\
  \bottomrule
\end{tabularx}
\end{table*}

\FloatBarrier

\subsection{Two-dimensional BEC: hBN and \texorpdfstring{$\alpha$-In$_2$Se$_3$}{alpha-In2Se3}}
\label{sec:two_dimensional_bec_results}

The acoustic sum rule requires equal and opposite B and N responses in hBN [Fig.~\ref{fig:two_dimensional_bec_structures}(a)]. The five-plane sequence of $\alpha$-In$_2$Se$_3$ distinguishes the In and Se sites used in the BEC comparison [Fig.~\ref{fig:two_dimensional_bec_structures}(b)]. ZStar also calculates vacuum levels and in-plane electrostatic-potential maps for two-dimensional materials (\ref{app:electrostatic_potential}).

\begin{center}
\begin{minipage}{\textwidth}
 \centering
 \includegraphics[width=0.72\textwidth]{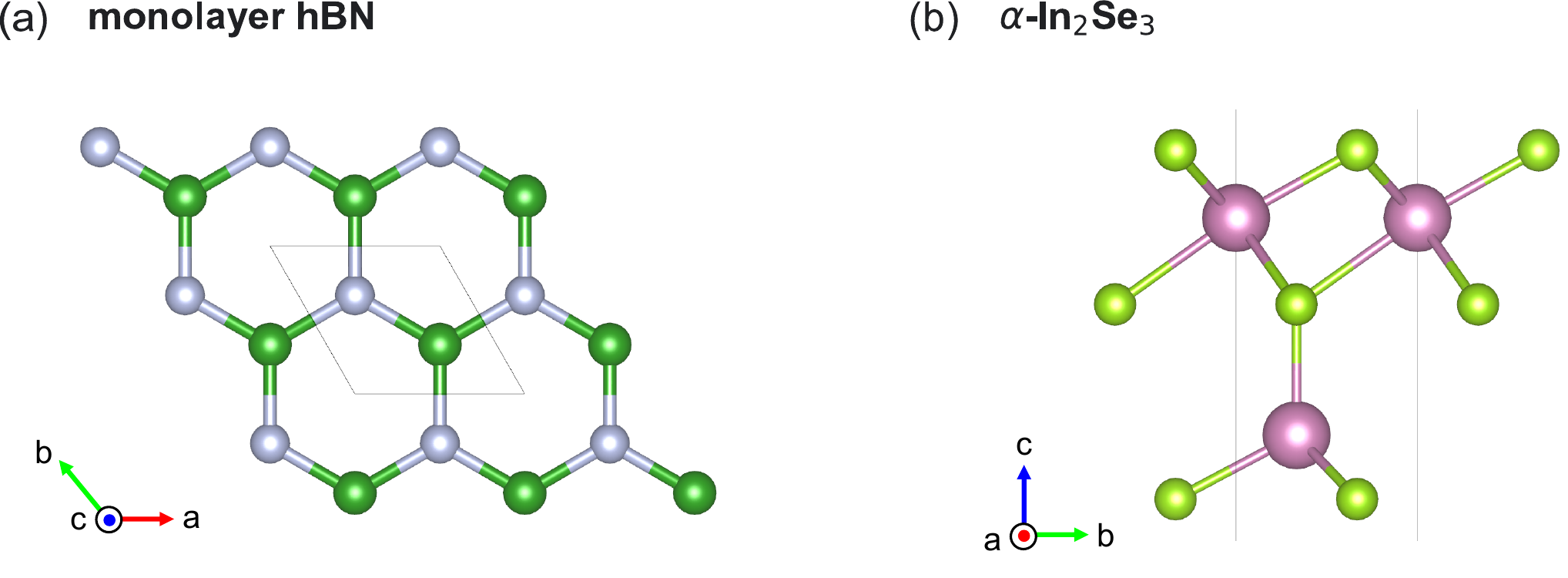}
 \captionsetup{hypcap=false}
 \captionof{figure}{Two-dimensional benchmark materials. (a) hBN. (b) Ferroelectric $\alpha$-In$_2$Se$_3$.}
 \label{fig:two_dimensional_bec_structures}
\end{minipage}
\end{center}

\FloatBarrier

For hBN, the hybrid route gives $Z^*_{\mathrm{B},\parallel}=2.702\,e$ from periodic Berry polarization and $Z^*_{\mathrm{B},z}=0.343\,e$ from the open-direction dipole. N carries the opposite tensor. The in-plane value matches the PBE/DFPT result of Sio and Giustino at the reported precision \cite{SioGiustino2022} and is close to the PZ-LDA result of Hu \textit{et al.}\ \cite{Hu2018hBN} (Table~\ref{tab:hbn_bec_literature}).

For hBN, the converged Berry-phase calculation gives normal BECs within $0.00020\,e$ of density integration on the same ABACUS outputs. Median single-core post-processing time per structure was 4.70~s for the hybrid route and 14.93~s for the all-direction Berry calculation.

For $\alpha$-In$_2$Se$_3$, the PBEsol-D3(0) calculation gives the site-resolved BECs in Table~\ref{tab:in2se3_bec_literature}. The in-plane values follow the pattern reported by Soleimani and Pourfath using PBEsol \cite{Soleimani2020}. The two In layers have distinct positive responses, and the central Se response is smaller in magnitude than those of the surface Se layers. The out-of-plane components are smaller than the in-plane components.

\begin{table}[!htbp]
 \centering
 \small
 \caption{BECs of two-dimensional hBN in units of $e$. Both sublattices are shown explicitly. For the two-atom primitive cell, the acoustic sum rule gives $Z^*_{\mathrm{N}}=-Z^*_{\mathrm{B}}$. A dash denotes a component not reported in the cited work.}
 \label{tab:hbn_bec_literature}
 \begin{tabularx}{\textwidth}{>{\raggedright\arraybackslash}p{0.20\textwidth}>{\raggedright\arraybackslash}p{0.20\textwidth}>{\centering\arraybackslash}p{0.10\textwidth}*{4}{>{\centering\arraybackslash}X}}
  \toprule
  Source & Software & XC & B$_{\parallel}$ & B$_z$ & N$_{\parallel}$ & N$_z$ \\
  \midrule
  ZStar & ABACUS+PYATB & PBE & 2.702 & 0.343 & $-2.702$ & $-0.343$ \\
  Sio and Giustino \cite{SioGiustino2022} & Quantum ESPRESSO & PBE & 2.702 & -- & $-2.702$ & -- \\
  Hu \textit{et al.}\ \cite{Hu2018hBN} & Quantum ESPRESSO & PZ-LDA & 2.697 & 0.281 & $-2.697$ & $-0.281$ \\
  \bottomrule
 \end{tabularx}
\end{table}

\FloatBarrier

\begin{table}[!htbp]
 \centering
 \small
 \caption{Layer-resolved BECs of two-dimensional $\alpha$-In$_2$Se$_3$ in units of $e$. Sites follow the sequence of $\mathrm{Se}(1)$--$\mathrm{In}(1)$--$\mathrm{Se}(0)$--$\mathrm{In}(2)$--$\mathrm{Se}(2)$ along $+z$.}
 \label{tab:in2se3_bec_literature}
 \footnotesize
 \begin{tabularx}{\textwidth}{>{\raggedright\arraybackslash}p{0.25\textwidth}>{\raggedright\arraybackslash}p{0.15\textwidth}>{\centering\arraybackslash}p{0.055\textwidth}*{5}{>{\centering\arraybackslash}X}}
  \toprule
  Source & Software & XC & Se(1) & In(1) & Se(0) & In(2) & Se(2) \\
  \midrule
  \multicolumn{8}{l}{\textit{In-plane response}, $Z^*_{\parallel}$} \\
  ZStar & ABACUS+PYATB & PBEsol & $-2.426$ & 2.639 & $-1.648$ & 3.864 & $-2.428$ \\

  Soleimani and Pourfath \cite{Soleimani2020} & VASP & PBEsol & $-2.505$ & 2.737 & $-1.697$ & 4.001 & $-2.535$ \\
  \midrule
  \multicolumn{8}{l}{\textit{Out-of-plane response}, $Z^*_{zz}$} \\
  ZStar & ABACUS+PYATB & PBEsol & $-0.219$ & 0.381 & $-0.341$ & 0.348 & $-0.169$ \\

  Soleimani and Pourfath \cite{Soleimani2020} & VASP & PBEsol & $-0.166$ & 0.289 & $-0.250$ & 0.255 & $-0.128$ \\
  \bottomrule
 \end{tabularx}
\end{table}

\FloatBarrier
\Needspace{7\baselineskip}
\subsection{One-dimensional BEC: BN(9,0) and \texorpdfstring{Sb$_2$S$_3$}{Sb2S3}}
\label{sec:one_dimensional_bec_results}

\begin{figure*}[htbp]
	\centering
\includegraphics[width=0.72\textwidth]{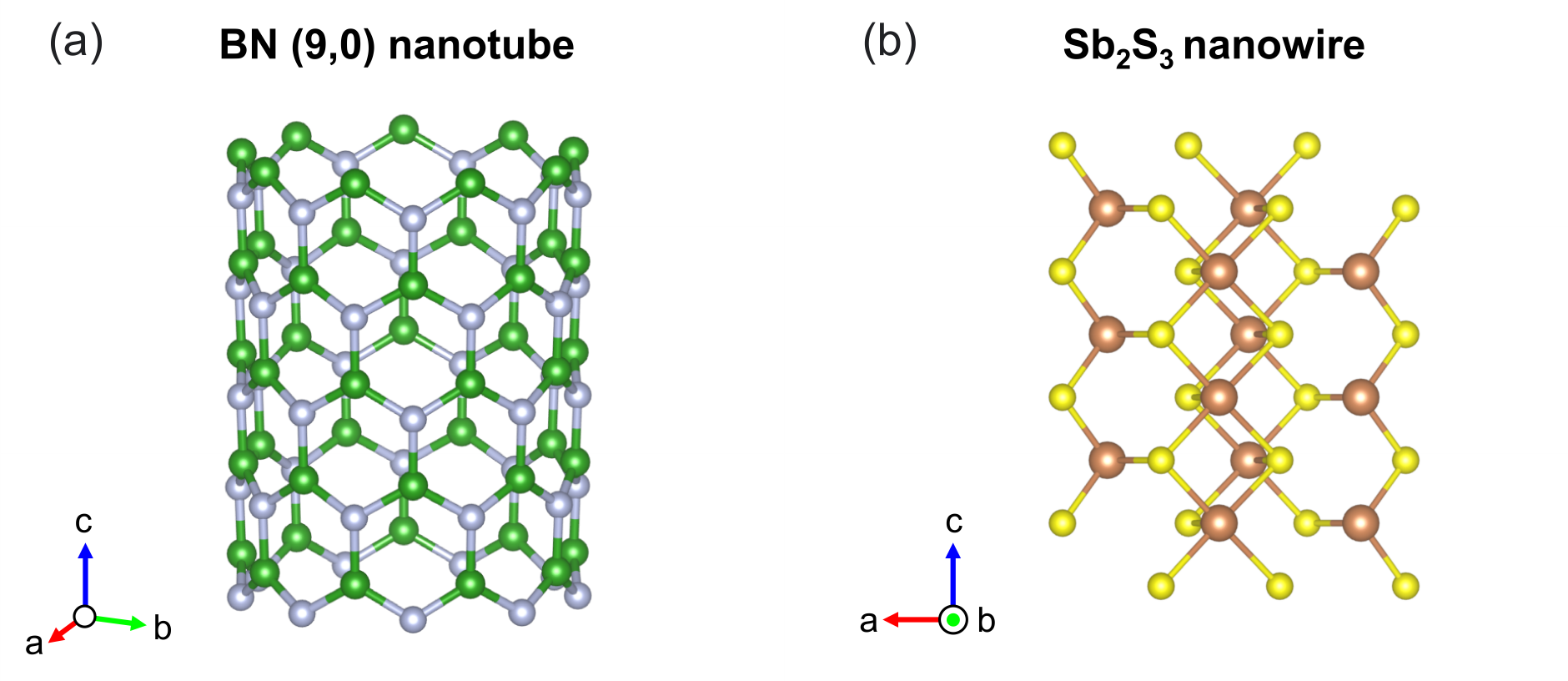}
	\caption{One-dimensional BEC examples: (a) the unpassivated BN(9,0) nanotube and (b) the isolated Sb$_2$S$_3$ chain. Both structures are centered in the transverse vacuum; $z$ is the periodic direction.}
	\label{fig:one_dimensional_bec_structures}
\end{figure*}

The one-dimensional examples extend the same unified response calculation to an unpassivated BN(9,0) nanotube [Fig.~\ref{fig:one_dimensional_bec_structures}(a)] and an isolated Sb$_2$S$_3$ chain [Fig.~\ref{fig:one_dimensional_bec_structures}(b)]. The optimized structures contain 36 and 10 atoms per axial repeat, respectively, and are centered in the transverse vacuum. BN(9,0) uses PBE and Sb$_2$S$_3$ uses PBE-D3(BJ); their optimized axial periods are 4.338 and 3.786~\AA{}, with band gaps of 3.813 and 1.501~eV. Both calculations use the hybrid convention of Section~\ref{sec:one_dimensional}: Berry-phase polarization along $z$ and real-space dipole derivatives along $x$ and $y$. The BEC and zone-center force constants are reconstructed from the same 56 and 20 displacement calculations, respectively, plus one undisplaced reference for each system.

For the two transverse directions of Sb$_2$S$_3$, the converged Berry-phase calculation and density integration agree within $0.001\,e$ on the same ABACUS outputs.

A fixed Cartesian component mixes radial and tangential responses at different azimuths. Table~\ref{tab:bn9_bec} therefore expresses each tensor in its atom-local orthonormal basis $(r,t,z)$ before averaging by chemical species. Guo \textit{et al.}~\cite{Guo2007BN} used the same local directions in their finite-field study with an isolated-tube transverse-field correction. We sum the full tensors in a common Cartesian frame to check charge neutrality. The archived tables list all nine Cartesian components for every atom and the range of each local diagonal component. The azimuthal ranges are below $0.006\,e$.

The VASP DFPT calculation uses PAW-PBE potentials \cite{Blochl1994PAW,Kresse1999PAW}, the same axial sampling and vacuum cell as ABACUS, and gives a band gap of 3.762~eV \cite{Kresse1996VASP}. Both calculations give the largest response along the tube axis. The axial BECs differ by 0.3\%, the tangential components by about 2\%, and the radial components by about $0.08\,e$. The VASP tensors have a maximum Cartesian charge-sum residual of $2\times10^{-5}\,e$, and the sum-rule projection changes the ABACUS components by at most $3.2\times10^{-4}\,e$.

\begin{table}[htbp]
	\centering\small
	\caption{Calculated BECs of the BN(9,0) nanotube, in units of $e$. Values are species averages after rotating each tensor into its local radial ($r$), tangential ($t$), and axial ($z$) basis.}
	\label{tab:bn9_bec}
	
	\begin{tabularx}{\textwidth}{
			llc
			*{6}{>{\centering\arraybackslash}X}
		}
		\toprule
		Source & Software & XC &
		$B_{rr}$ & $B_{tt}$ & $B_{zz}$ &
		$N_{rr}$ & $N_{tt}$ & $N_{zz}$ \\
		\midrule
		ZStar & ABACUS+PYATB & PBE &
		0.397 & 1.256 & 2.745 &
		-0.474 & -1.178 & -2.745 \\
		ZStar & VASP & PBE &
		0.318 & 1.234 & 2.754 &
		-0.398 & -1.154 & -2.754 \\
		\bottomrule
	\end{tabularx}
\end{table}

\begin{table}[htbp]
\centering\footnotesize
\caption{Diagonal Born effective charges of the isolated Sb$_2$S$_3$ chain, in units of $e$. Each site represents two symmetry-equivalent atoms. The VASP finite-displacement calculation uses the VASP-relaxed structure. The CRYSTAL axes and sites follow the ZStar convention.}
\label{tab:sb2s3_bec}
\begin{tabularx}{\textwidth}{@{}lll*{5}{>{\centering\arraybackslash}X}@{}}
\toprule
Source & Software & XC & Sb(1) & Sb(2) & S(1) & S(2) & S(3) \\
\midrule
\multicolumn{8}{l}{$Z^*_{xx}$: transverse response} \\
ZStar & ABACUS+PYATB & PBE-D3(BJ) & 1.100 & 0.867 & $-0.530$ & $-0.756$ & $-0.681$ \\
ZStar & VASP (finite disp.) & PBE-D3(BJ) & $1.168$ & $0.995$ & $-0.525$ & $-0.867$ & $-0.770$ \\
Ulian \cite{Ulian2026Sb2S3Data} & CRYSTAL & B3LYP-D3(BJ) & 1.143 & 0.999 & $-0.514$ & $-0.859$ & $-0.768$ \\
\midrule
\multicolumn{8}{l}{$Z^*_{yy}$: transverse response} \\
ZStar & ABACUS+PYATB & PBE-D3(BJ) & 0.313 & 0.588 & $-0.164$ & $-0.244$ & $-0.493$ \\
ZStar & VASP (finite disp.) & PBE-D3(BJ) & $0.323$ & $0.629$ & $-0.175$ & $-0.239$ & $-0.539$ \\
Ulian \cite{Ulian2026Sb2S3Data} & CRYSTAL & B3LYP-D3(BJ) & 0.367 & 0.670 & $-0.199$ & $-0.276$ & $-0.561$ \\
\midrule
\multicolumn{8}{l}{$Z^*_{zz}$: axial response} \\
ZStar & ABACUS+PYATB & PBE-D3(BJ) & 4.642 & 5.972 & $-4.101$ & $-3.353$ & $-3.159$ \\
ZStar & VASP (finite disp.) & PBE-D3(BJ) & $4.609$ & $6.046$ & $-4.085$ & $-3.399$ & $-3.171$ \\
Ulian \cite{Ulian2026Sb2S3Data} & CRYSTAL & B3LYP-D3(BJ) & 4.341 & 6.135 & $-3.786$ & $-3.430$ & $-3.260$ \\
\bottomrule
\end{tabularx}
\end{table}

For Sb$_2$S$_3$, we transform the reference axes as $(x,y,z)_{\mathrm{ZStar}}=(y,z,x)_{\mathrm{CRYSTAL}}$ and match sites by their coordinates. The VASP BECs come from central differences of the total dipole along twelve phonon-mode displacements. Inversion, mirror symmetry, and the acoustic sum rule map the mode derivatives to atomic tensors.

\FloatBarrier
\Needspace{8\baselineskip}
\subsection{Molecular atomic polar tensors: \texorpdfstring{H$_2$O and CH$_4$}{H2O and CH4}}

We compare generalized atomic polar tensor (APT) charges for polar H$_2$O [Fig.~\ref{fig:molecular_apt_structures}(a)] and nonpolar tetrahedral CH$_4$ [Fig.~\ref{fig:molecular_apt_structures}(b)]. Water has larger generalized APT charge magnitudes. Methane has small APT traces and anisotropic hydrogen tensors.

\begin{center}
\begin{minipage}{\textwidth}
 \centering
 \includegraphics[width=0.72\textwidth]{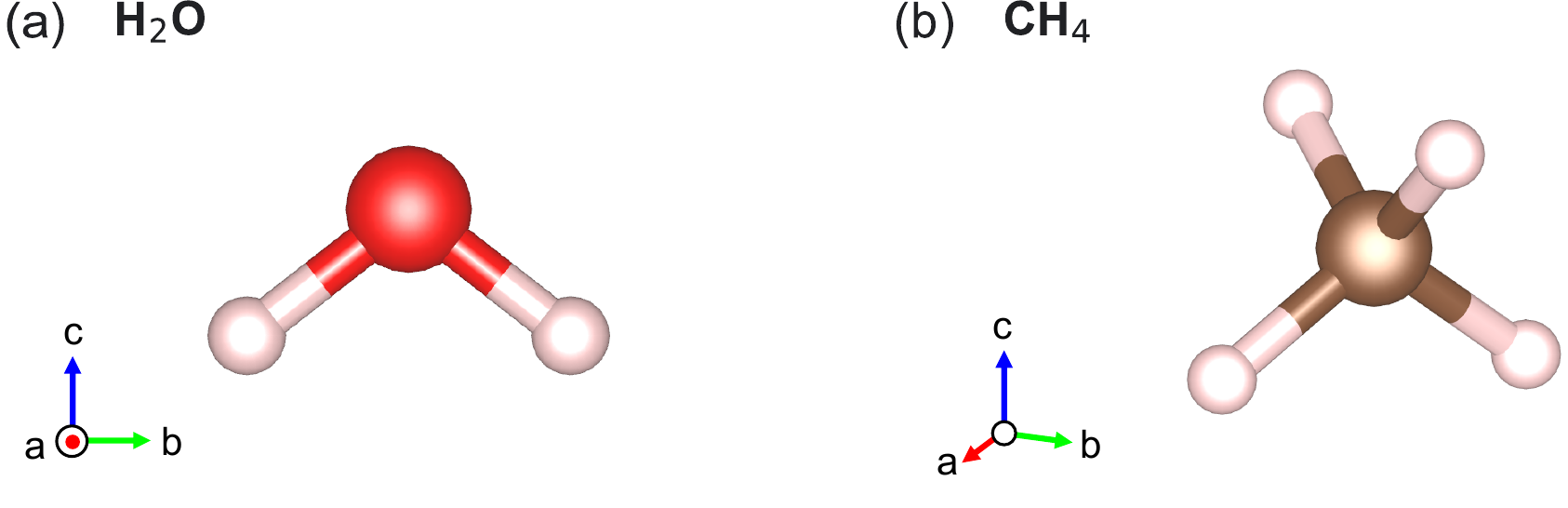}
 \captionof{figure}{Molecular benchmark structures. (a) Polar H$_2$O. (b) Tetrahedral and nonpolar CH$_4$.}
 \label{fig:molecular_apt_structures}
\end{minipage}
\end{center}

For CH$_4$, occupied position-matrix elements and charge-density integration give BECs within $0.00136\,e$ at the same HSE reference and three displaced geometries, before imposing charge neutrality. For the PBE reference and three displacements in a $20$~\AA{} cell, the $8^3$ and $16^3$ Berry meshes give the same dipole changes at the output precision.

We compare the rotationally invariant generalized APT charge $q^{\mathrm{GAPT}}_\kappa=\mathrm{Tr}(\boldsymbol{A}_\kappa)/3$. Both molecular calculations use ABACUS. The PBE and HSE results for H$_2$O follow the same charge pattern as the Hartree--Fock values reported by \AA strand \textit{et al.} \cite{Astrand1998Water}.  Table~\ref{tab:h2o_apt_literature} compare the results of H$_2$O. Full tensors and HSE details are given in \ref{app:molecular_tensors}.

 For tetrahedral CH$_4$, the PBE values are $-0.021\,e$ for C and $+0.005\,e$ for H, while the HSE calculation gives $-0.015\,e$ and $+0.004\,e$. Their molecular sums are zero before rounding after symmetry reconstruction. The small APT traces coexist with finite anisotropic H response components that generate the infrared-active $T_2$ modes. Reported GAPT values change sign with the force field, basis set, and electronic-structure method because the trace is close to zero \cite{FerreiraBassi1987Methane,Oliveira2000GAPT,Richter2021GAPT}. Table~\ref{tab:ch4_apt_literature} compares the traces with the B3LYP reference.

\begin{table*}[htbp]
 \centering
 \small
 \caption{Generalized APT charges of isolated H$_2$O in units of $e$. Each charge is one third of the APT trace.}
 \label{tab:h2o_apt_literature}
 \begin{tabularx}{\textwidth}{>{\raggedright\arraybackslash}p{0.22\textwidth}>{\raggedright\arraybackslash}p{0.22\textwidth}>{\raggedright\arraybackslash}p{0.24\textwidth}*{2}{>{\centering\arraybackslash}X}}
  \toprule
  Source & Software & XC/method & $q^{\mathrm{GAPT}}_{\mathrm O}$ & $q^{\mathrm{GAPT}}_{\mathrm H}$ \\
  \midrule
  ZStar & ABACUS & PBE & $-0.481$ & $+0.240$ \\
 ZStar & ABACUS & HSE & $-0.506$ & $+0.253$ \\

  \AA strand \textit{et al.}\ \cite{Astrand1998Water} & response calculation & HF/aug-cc-pV5Z & $-0.564$ & $+0.282$ \\
  \bottomrule
 \end{tabularx}
\end{table*}

\begin{table*}[htbp]
 \centering
 \small
 \caption{Generalized APT charges of isolated CH$_4$ in units of $e$, compared with the B3LYP/6-31+G(d,p) calculation of Oliveira \textit{et al.} Each charge is one third of the APT trace. Molecular neutrality gives $q_{\mathrm C}+4q_{\mathrm H}=0$.}
 \label{tab:ch4_apt_literature}
 \begin{tabularx}{\textwidth}{>{\raggedright\arraybackslash}p{0.22\textwidth}>{\raggedright\arraybackslash}p{0.20\textwidth}>{\raggedright\arraybackslash}p{0.26\textwidth}*{2}{>{\centering\arraybackslash}X}}
  \toprule
  Source & Software & XC/method & $q^{\mathrm{GAPT}}_{\mathrm C}$ & $q^{\mathrm{GAPT}}_{\mathrm H}$ \\
  \midrule
  ZStar & ABACUS & PBE & $-0.021$ & $+0.005$ \\
  ZStar & ABACUS & HSE & $-0.015$ & $+0.004$ \\
  Oliveira \textit{et al.}\ \cite{Oliveira2000GAPT} & Gaussian & B3LYP/6-31+G(d,p) & $-0.024$ & $+0.006$ \\
  \bottomrule
 \end{tabularx}
\end{table*}

\FloatBarrier
\subsection{Static and frequency-dependent dielectric response}
\label{sec:dielectric_response}

We calculate the dielectric response of tetragonal HfO$_2$ and the sheet polarizability of monolayer hBN. The electronic contribution is combined with the phonon contribution obtained from the BECs and zone-center modes.

For tetragonal HfO$_2$, the calculation with PBEsol functional gives
\begin{equation}
 \boldsymbol{\epsilon}^{\infty}
 =\begin{pmatrix}5.162&0&0\\0&5.162&0\\0&0&4.780\end{pmatrix},
 \qquad
 \boldsymbol{\epsilon}^{\mathrm{ph}}
 =\begin{pmatrix}70.599&0&0\\0&70.599&0\\0&0&13.265\end{pmatrix},
\end{equation}
\begin{equation}
 \boldsymbol{\epsilon}(0)
 =\boldsymbol{\epsilon}^{\infty}+\boldsymbol{\epsilon}^{\mathrm{ph}}
 =\begin{pmatrix}75.761&0&0\\0&75.761&0\\0&0&18.045\end{pmatrix}.
\end{equation}
The low-frequency $E_u$ pair at 96.1~cm$^{-1}$ gives the strongest in-plane peak and contributes 92.6\% of the in-plane static phonon response. The $A_{2u}$ mode gives the out-of-plane resonance. The large in-plane response from the soft polar mode is consistent with the mechanism identified for tetragonal HfO$_2$ by He \textit{et al.}\ \cite{He2026HfO2Dielectric}.
\begin{figure*}[htbp]
	\centering
\includegraphics[width=0.76\textwidth]{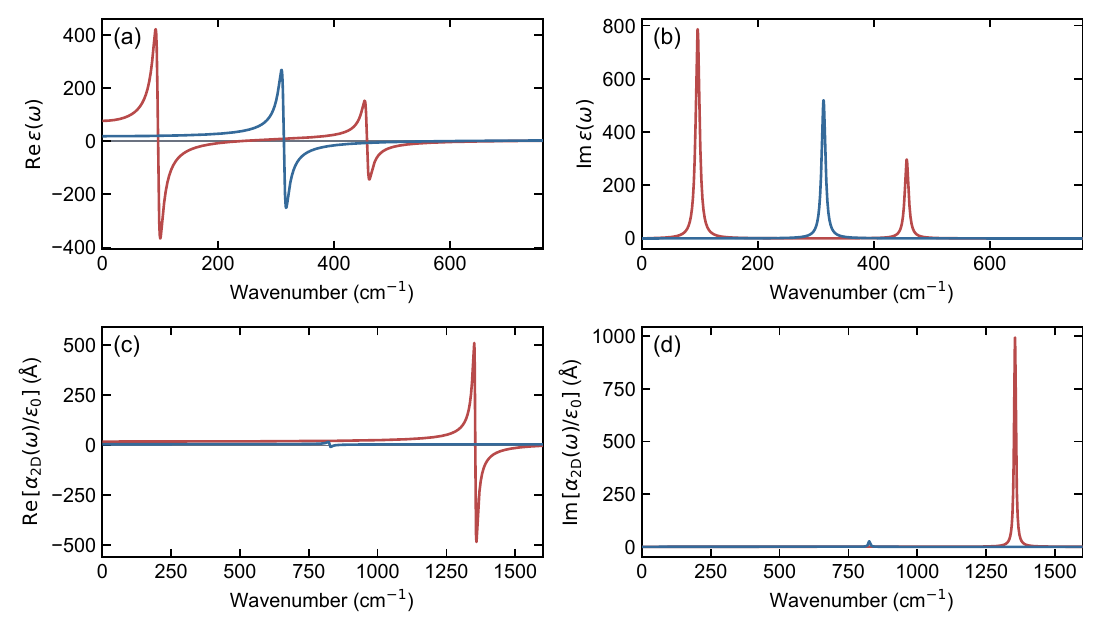}
	\caption{Frequency-dependent dielectric response of tetragonal HfO$_2$ (a,b) and monolayer hBN (c,d). The left and right panels show the real and imaginary parts, respectively. Red and blue denote the in-plane ($xx=yy$) and out-of-plane ($zz$) components, respectively. The Lorentzian damping is 8~cm$^{-1}$.}
	\label{fig:dielectric_response_examples}
\end{figure*}
\FloatBarrier

\Needspace{5\baselineskip}
The sheet polarizability relates the dipole moment per unit area to the applied field \cite{Laturia2018}. For monolayer hBN, its in-plane and out-of-plane static components divided by $\varepsilon_0$ are 18.026 and 4.545~\AA{}, respectively. The $E'$ mode gives the dominant in-plane resonance at 1354.86~cm$^{-1}$, and the $A_2''$ mode gives the weaker out-of-plane resonance at 825.21~cm$^{-1}$. Dai \textit{et al.}\ reported 1362.4~cm$^{-1}$ for the in-plane transverse-optical mode \cite{Dai2019hBN}.

\FloatBarrier
\Needspace{8\baselineskip}
\subsection{Piezoelectric response of insulating bulk crystals}
\label{sec:piezoelectric_results}

We calculate the relaxed-ion piezoelectric coefficients of wurtzite AlN and ZnO ($P6_3mc$) using the finite-strain method in Section~\ref{sec:piezoelectric_theory}. Both crystals have three independent coefficients, $e_{31}$, $e_{33}$, and $e_{15}$, when $z$ follows the polar $c$ axis. The ABACUS+PYATB route samples all six engineering strains at $\pm0.5\%$ plus the unstrained reference, relaxes internal coordinates at fixed strained cells, and fits the actual written strains. The reported $\mathbf e$ tensors are proper relaxed-ion responses, and $\mathbf d$ is obtained from the corresponding fixed-field elastic matrix.

Table~\ref{tab:piezo_aln} compares the AlN calculations with the piezoelectric stress coefficients of de Jong \textit{et al.}\ \cite{deJong2015Piezo} and the strain coefficients obtained by Kim \textit{et al.}\ from five resonance modes of room-temperature AlN single crystals \cite{Kim2015AlNPiezo}. ABACUS/PBE gives $e_{31}=-0.569$, $e_{33}=1.388$, and $e_{15}=-0.297$~C/m$^2$. The differences in magnitude from the de Jong results are 1.9\%, 5.0\%, and 2.8\%, respectively. Both tensors follow wurtzite symmetry and give the same sign for each coefficient. The zero-temperature AlN calculations and room-temperature resonance measurements give piezoelectric strain coefficients of the same sign and similar magnitude. The largest relative differences occur for $d_{15}$. Shear compliance enters the conversion from the piezoelectric stress coefficients to $\mathbf d$. The native VASP $\mathbf e$ tensor differs from the dataset of de Jong \textit{et al.}\ \cite{deJong2015Piezo} by 1.67\% in Frobenius norm. The symmetry-forbidden VASP components are below $4\times10^{-5}$~C/m$^2$. The independently optimized ABACUS and VASP models differ by 4.47\%, 1.32\%, and 5.98\% in their complete $\mathbf e$, $\mathbf C^E$, and $\mathbf d$ tensors, respectively. The positive-definite native VASP elastic matrix gives $d_{33}=5.324$~pm/V, and the product $\mathbf d\mathbf C^E$ recovers $\mathbf e$. In the VASP decomposition for AlN [Eq.~(\ref{eq:piezo_internal_relaxation})], the clamped-ion term is $-0.421$~C/m$^2$ and the ionic term is $1.883$~C/m$^2$, giving a total $e_{33}$ of $1.461$~C/m$^2$. Internal relaxation reverses the sign and increases the magnitude of the longitudinal response. The same calculation gives the dielectric tensor at fixed strain. 

\begin{table*}[!htbp]
\centering\footnotesize
\setlength{\tabcolsep}{4.5pt}
\caption{Proper, relaxed-ion piezoelectric response of wurtzite AlN in the $3\times6$ convention. The $e$ coefficients are in C/m$^2$ and the $d$ coefficients are in pm/V; dashes denote unreported quantities.}
\label{tab:piezo_aln}
\begin{tabularx}{\textwidth}{>{\raggedright\arraybackslash}p{0.16\textwidth}>{\raggedright\arraybackslash}p{0.14\textwidth}>{\centering\arraybackslash}p{0.08\textwidth}*{6}{>{\centering\arraybackslash}X}}
\toprule
Source & Software/method & XC/model & $e_{31}$ & $e_{33}$ & $e_{15}$ & $d_{31}$ & $d_{33}$ & $d_{15}$ \\
\midrule
ZStar & ABACUS+PYATB & PBE & $-0.569$ & $1.388$ & $-0.297$ & $-2.068$ & $4.975$ & $-2.628$ \\
ZStar & VASP & PBE & $-0.581$ & $1.461$ & $-0.310$ & $-2.190$ & $5.324$ & $-2.768$ \\
de Jong \textit{et al.}\ \cite{deJong2015Piezo} & VASP & PBE & $-0.580$ & $1.461$ & $-0.289$ & --- & --- & --- \\
Kim \textit{et al.}\ \cite{Kim2015AlNPiezo} & resonance & experiment & --- & --- & --- & $-1.9$ & $6.5$ & $-1.9$ \\
\bottomrule
\end{tabularx}
\end{table*}
\FloatBarrier

The ZnO comparison also uses PBE (Table~\ref{tab:piezo_zno}). The ABACUS+PYATB values of $e_{31}$, $e_{33}$, and $e_{15}$ differ in magnitude from the de Jong PBE data by 3.0\%, 1.0\%, and 0.3\%, respectively; the corresponding native VASP differences are 0.1\%, 0.5\%, and 3.7\%. Both routes retain the expected $6mm$ tensor form. Kobiakov gives strain coefficients from room-temperature resonance measurements \cite{Kobiakov1980ZnOPiezo}. ZnO has greater shear compliance than AlN. Both calculations reproduce the signs and relative magnitudes of the measured $d$ coefficients. The complete ABACUS and VASP tensors differ by 2.16\% for $\mathbf e$, 4.43\% for $\mathbf C^E$, and 11.85\% for $\mathbf d$, despite the similar $d_{33}$ values. Both elastic matrices are positive definite, and both sets of tensors satisfy $\mathbf e=\mathbf d\mathbf C^E$.

\begin{table*}[!htbp]
\centering\footnotesize
\setlength{\tabcolsep}{4.5pt}
\caption{Proper, relaxed-ion piezoelectric response of wurtzite ZnO in the $3\times6$ convention. The $e$ coefficients are in C/m$^2$ and the $d$ coefficients are in pm/V; dashes denote unreported quantities.}
\label{tab:piezo_zno}
\begin{tabularx}{\textwidth}{>{\raggedright\arraybackslash}p{0.16\textwidth}>{\raggedright\arraybackslash}p{0.14\textwidth}>{\centering\arraybackslash}p{0.08\textwidth}*{6}{>{\centering\arraybackslash}X}}
\toprule
Source & Software/method & XC/model & $e_{31}$ & $e_{33}$ & $e_{15}$ & $d_{31}$ & $d_{33}$ & $d_{15}$ \\
\midrule
ZStar & ABACUS+PYATB & PBE & $-0.522$ & $1.047$ & $-0.384$ & $-4.669$ & $9.666$ & $-9.217$ \\
ZStar & VASP & PBE & $-0.537$ & $1.042$ & $-0.399$ & $-4.887$ & $9.750$ & $-10.828$ \\
de Jong \textit{et al.}\ \cite{deJong2015Piezo} & VASP & PBE & $-0.538$ & $1.037$ & $-0.385$ & --- & --- & --- \\

Kobiakov \cite{Kobiakov1980ZnOPiezo} & resonance & experiment & --- & --- & --- & $-5.12$ & $12.3$ & $-8.3$ \\
\bottomrule
\end{tabularx}
\end{table*}

\FloatBarrier
\subsection{Phonon dispersion and LO--TO splitting with the non-analytic correction}
\label{sec:phonon_nac_results}

For cubic $Pm\bar{3}m$ BaTiO$_3$, we calculate the dispersion from forces in a
$2\times2\times2$ Phonopy supercell. The non-analytic term uses the BEC and
electronic dielectric tensors from the bulk calculation. The high-symmetry path
is generated from the standardized structure using Seekpath and spglib
\cite{Hinuma2017Seekpath,Togo2024Spglib}. The
workflow calculates the dispersion and DOS with and without the non-analytic
correction. The example archive includes both the dispersion and DOS.

Figure~\ref{fig:phonon_nac_bto} compares the two dispersions. The blue curves show
the analytic force-constant result without NAC. The red curves include
the direction-dependent dipole--dipole term from Eq.~(\ref{eq:nac_term}).
Near
 $\Gamma$, the macroscopic field shifts the polar branches, producing
 the LO--TO separation, while nonpolar branches
 remain essentially unchanged
\cite{Zhong1994,MalenfantThuot2024}.

\begin{center}
\begin{minipage}{\textwidth}
 \centering
 \includegraphics[width=0.65\textwidth]{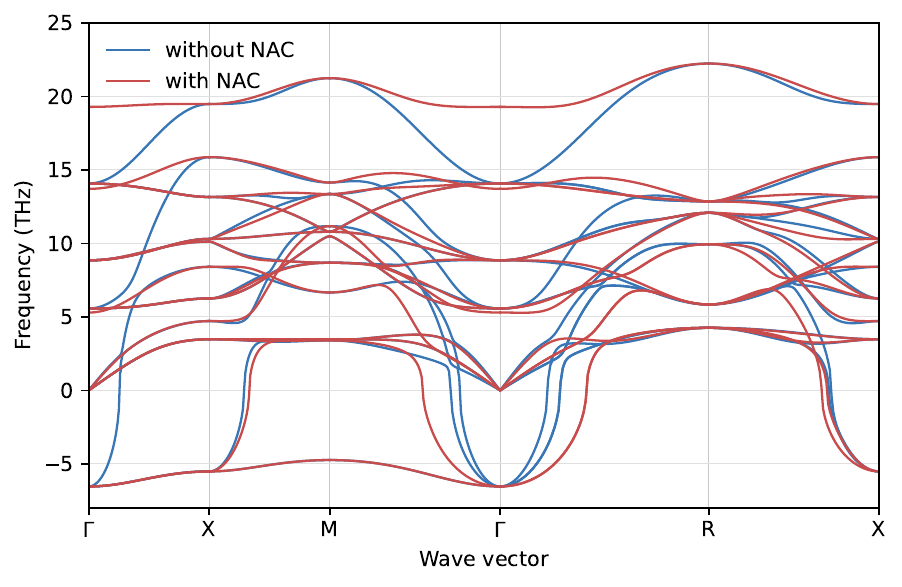}
 \captionsetup{hypcap=false}
 \captionof{figure}{Phonon dispersion of cubic BaTiO$_3$ with and without the
 non-analytic correction. The blue curves use the short-range force constants. The red curves include the long-range dipole term.}
 \label{fig:phonon_nac_bto}
\end{minipage}
\end{center}

\FloatBarrier
\subsection{Compatibility with field-aware machine-learning force fields}
\label{sec:field_aware_ml}

Field-aware interatomic potentials can learn polarization or dynamic charges alongside atomic energies and forces. Recently, learned electrical responses have been incorporated into atomistic simulations using models such as Allegro-pol \cite{Musaelian2023Allegro,Falletta2025}, MACE-Field \cite{Martin2026MACEField}, DREAM--Allegro \cite{Deng2026DREAM}, HotPP \cite{Wang2024HotPP,Guo2026HotPPPME}, and qNEP \cite{Fan2026qNEP}. ZStar provides structures and first-principles response data for training these models. Here we use NextForce, a field-aware force field developed by the authors, to illustrate the workflow.

\begin{figure*}[!t]
 \centering
 \includegraphics[width=0.74\textwidth]{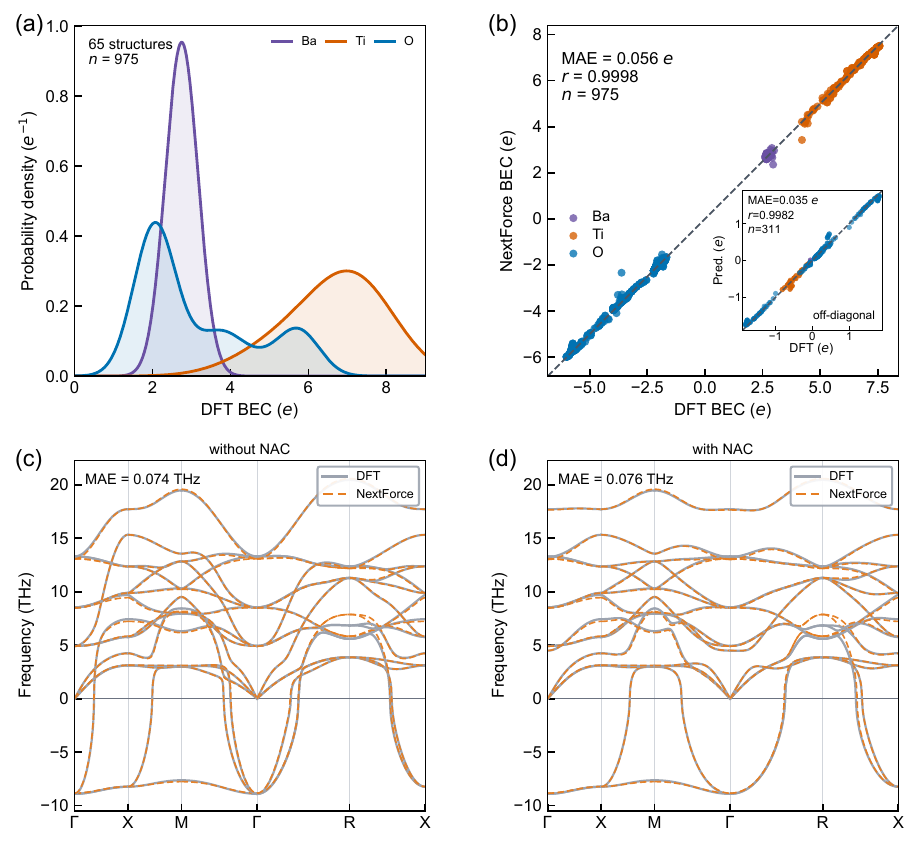}
 \caption{Compatibility benchmark with NextForce.
 (a) Element-resolved kernel-density estimates of the absolute values of all
 975 diagonal DFT BEC components in the 65-parent reference subset. (b)
 DFT--NextForce parity for
 all 975 diagonal BEC components of those structures. The inset isolates the
 311 nontrivial off-diagonal components satisfying
 $|Z^*_{\alpha\beta,\mathrm{DFT}}|>0.05e$.
  (c,d) DFT (gray solid) and NextForce (orange
 dashed) phonon
 dispersions from the same cubic BaTiO$_3$ structure, (c) without and (d) with NAC.
NextForce and DFT results share the same electronic dielectric constant  $\epsilon_\infty=6.87$ for both curves. All results use the same
 frozen model.}
 \label{fig:nextforce_validation}
\end{figure*}

The BaTiO$_3$ training, validation, and test sets contain 469, 114, and 87 structures. All have mechanical labels. Of the 469 training structures, 87 (18.6\%) also have Berry-phase polarization labels, compared modulo the full Cartesian polarization-quantum lattice. We train and select the model using energy, force, stress, and polarization. The BECs are derivatives of the learned dipole surface. We also evaluated BECs for two tetragonal $P4mm$ parent structures after fixing the model and evaluation criteria. On the zero-field test set, the energy, force, stress, and polarization MAEs are $0.188$~meV/atom, $0.01$~eV/\AA{}, $0.355$~meV/\AA$^3$, and $1.40\times10^{-3}$~C/m$^2$, respectively.

The BEC reference set contains 65 primitive parents (20 cubic, 19 tetragonal, 14 orthorhombic, and 12 rhombohedral), with 45, 10, and 10 in the mechanical training, validation, and test sets. Figure~\ref{fig:nextforce_validation}(a) shows the diagonal-component magnitudes by element. Across all three splits, the 975 diagonal components in Fig.~\ref{fig:nextforce_validation}(b) give an MAE of $0.0561~e$ and a Pearson coefficient of $0.9998$. The 311 off-diagonal components selected for the inset give an MAE of $0.0353~e$. Over all 1950 off-diagonal components, the MAE is $0.00658~e$, and 74.8\% of the DFT entries are zero. The held-out two-parent $P4mm$ check gives MAEs of $0.0330~e$ over all components and $0.0989~e$ over the diagonal components.

The same frozen model gives phonon frequency MAEs of $0.074$~THz without NAC and $0.076$~THz with NAC [Fig.~\ref{fig:nextforce_validation}(c,d)] compared with the DFT results in Fig.~\ref{fig:phonon_nac_bto}. Agreement extends to the low-frequency branch pattern near $\Gamma$. The dispersion without NAC uses the learned force constants. The corrected dispersion also uses BECs derived from the learned dipole surface and shares the electronic dielectric constant with the DFT calculation.

\FloatBarrier
\subsection{Infrared and Raman spectra}
\label{sec:spectroscopy_results}

\begin{figure*}[!t]
 \centering
 \newcommand{\FigureTenLegend}[4]{%
   \put(#1,#2){\makebox(0,0)[b]{\sffamily
     \fontsize{#3\unitlength}{#3\unitlength}\selectfont #4}}}
 \newcommand{\FigureTenLine}[1]{\rule{#1em}{.12em}}
 \newcommand{\FigureTenSolid}{\FigureTenLine{1.45}}
 \newcommand{\FigureTenDashed}{\FigureTenLine{.54}\kern.24em\FigureTenLine{.54}\kern.13em}
 \newcommand{\FigureTenEntry}[3]{%
   \raisebox{.27em}{\textcolor[HTML]{#1}{\csname FigureTenStyle#2\endcsname}}\kern.28em#3}
 \let\FigureTenStylesolid\FigureTenSolid
 \let\FigureTenStyledashed\FigureTenDashed
 \begin{overpic}[width=0.94\textwidth,permil]{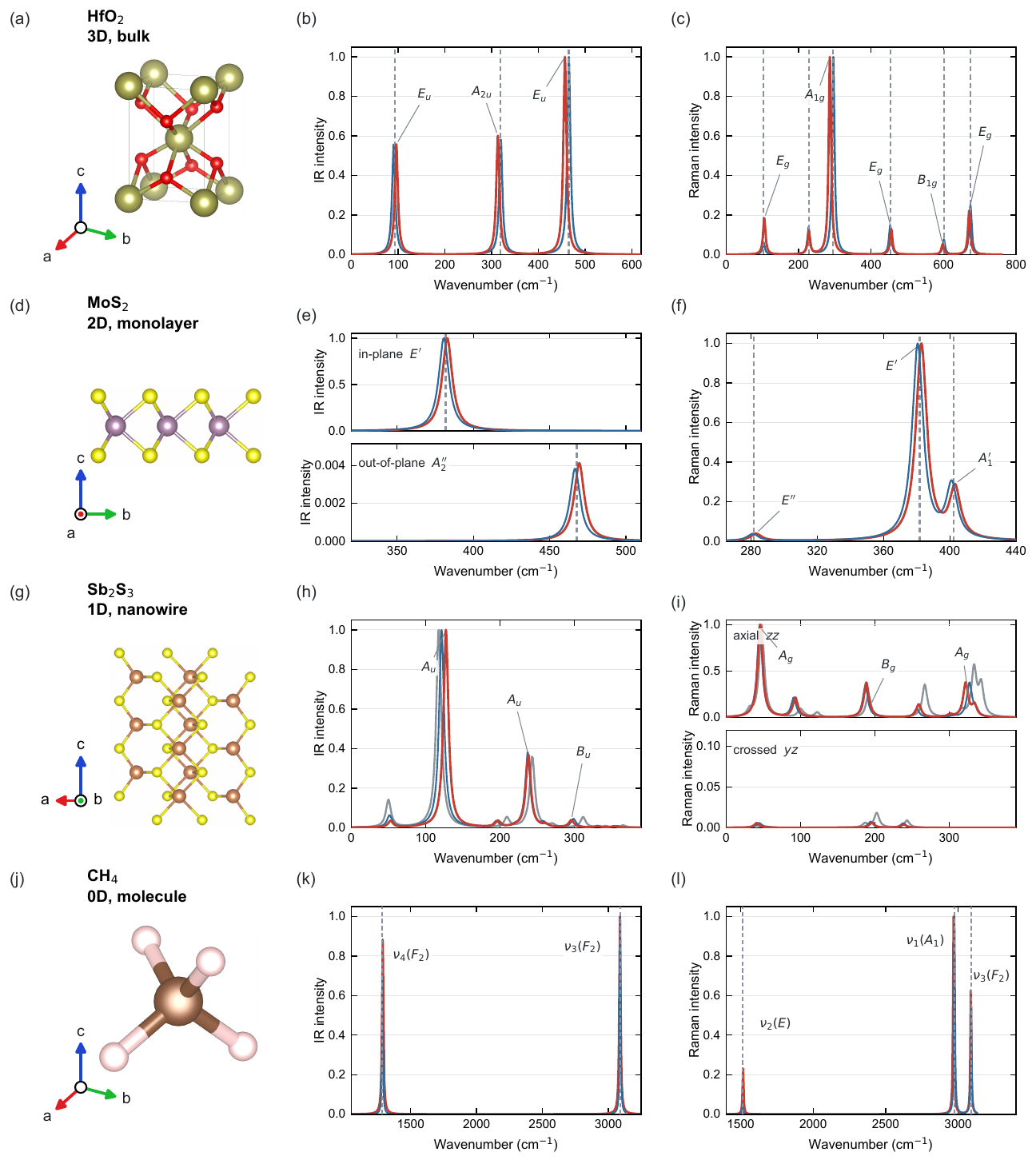}
   \input{Figure_10_citations.tex}
 \end{overpic}
\caption{Structures and vibrational spectra across four dimensionalities. (a) Tetragonal HfO$_2$ structure, (b) IR spectra, and (c) Raman spectra. (d) Monolayer MoS$_2$ structure, (e) in-plane and out-of-plane IR spectra (top and bottom), and (f) Raman spectra summed over in-plane and mixed $xz/yz$ channels. (g) Isolated Sb$_2$S$_3$ chain, (h) IR spectra, and (i) axial $zz$ and crossed $yz$ Raman spectra (top and bottom). (j) CH$_4$ structure, (k) IR spectra, and (l) Raman spectra. Red and blue curves are ZStar results produced by ABACUS and VASP, respectively. Gray dashed lines and curves show reference results.}
 \label{fig:cross_dimensional_spectroscopy}
\end{figure*}

Figure~\ref{fig:cross_dimensional_spectroscopy} compares infrared and Raman spectra for tetragonal HfO$_2$, monolayer MoS$_2$, a single Sb$_2$S$_3$ chain, and CH$_4$. We compare the ABACUS and VASP results with published frequencies or spectra to assess the mode assignments and peak positions in each dimensionality. The literature frequencies are shown as gray markers, and the published Sb$_2$S$_3$ spectra are shown as gray curves.

\Needspace{5\baselineskip}
In tetragonal HfO$_2$, the centrosymmetric $D_{4h}$ point group separates IR- and Raman-active modes [Fig.~\ref{fig:cross_dimensional_spectroscopy}(a--c)]. The three IR peaks from ABACUS at 96.1, 313.3, and 456.8~cm$^{-1}$ and from VASP at 90.3, 319.5, and 465.1~cm$^{-1}$ correspond to the Phonix $E_u$, $A_{2u}$, and $E_u$ frequencies at 92.4, 319.5, and 464.9~cm$^{-1}$ \cite{Ohnishi2026Phonix}. The dominant $A_{1g}$ Raman peak occurs at 286.2~cm$^{-1}$ in ABACUS and 295.6~cm$^{-1}$ in VASP, around the Phonix frequency of 293.8~cm$^{-1}$. Both calculations use PBEsol and their respective relaxed geometries.

Monolayer MoS$_2$ in Fig.~\ref{fig:cross_dimensional_spectroscopy}(d--f) has $D_{3h}$ symmetry. The $E'$ mode is active in both in-plane IR and Raman spectra. The $A_2''$ mode gives out-of-plane IR, $A_1'$ gives in-plane Raman scattering, and $E''$ appears in mixed $xz$ or $yz$ Raman polarizations. On the same monolayer structure with PBE-D3(BJ), ABACUS gives 282.97, 382.83, 403.12, and 469.31~cm$^{-1}$, and VASP gives 281.02, 380.59, 401.13, and 466.60~cm$^{-1}$, in the order $E''$, $E'$, $A_1'$, $A_2''$. The two codes differ by 2.22~cm$^{-1}$ on average and by at most 2.72~cm$^{-1}$. The corresponding PBE-D3 values from Table~3 of Ulian and Valdr\`e\ are 281.9, 381.8, 402.3, and 467.7~cm$^{-1}$ \cite{Ulian2023MoS2}. The largest deviations from those values are 1.61~cm$^{-1}$ for ABACUS and 1.21~cm$^{-1}$ for VASP. 
In each calculation the out-of-plane IR peak reaches about 0.4\% of the corresponding in-plane IR peak. In the static independent-particle Raman spectra, the mixed-polarization $E''$ peak reaches 3.67\% and 3.60\% of the corresponding in-plane Raman peak in ABACUS and VASP, respectively.

For the isolated Sb$_2$S$_3$ chain [Fig.~\ref{fig:cross_dimensional_spectroscopy}(g--i)], the two largest IR peaks appear near 127 and 238~cm$^{-1}$ in ABACUS, 121 and 237~cm$^{-1}$ in VASP, and 118 and 243~cm$^{-1}$ in CRYSTAL \cite{Ulian2026Sb2S3Data}. The VASP spectrum uses the finite-displacement BECs in Table~\ref{tab:sb2s3_bec}. The axial $zz$ Raman spectra show strong peaks at low frequency and around 190 and 320--335~cm$^{-1}$. The CRYSTAL spectrum also shows a separate feature near 267~cm$^{-1}$. The maximum crossed $yz$ response is 0.64\%, 0.76\%, and 1.84\% of the corresponding axial peak in ABACUS, VASP, and CRYSTAL, respectively.

In tetrahedral CH$_4$ [Fig.~\ref{fig:cross_dimensional_spectroscopy}(j--l)], the $T_2$ modes $\nu_4$ and $\nu_3$ appear in both IR and Raman spectra. The $\nu_2(E)$ and $\nu_1(A_1)$ modes are Raman active only. The figure uses $F_2$ for $T_2$. The gas-phase VASP/PBE reference frequencies of Yadavalli \textit{et al.}\ are 1285.7, 1511.5, 2975.0, and 3091.3~cm$^{-1}$ \cite{Yadavalli2021}. The four peaks in our VASP spectra fall within about 2.1~cm$^{-1}$ of those markers.

The ABACUS and VASP spectra reproduce the same principal IR and Raman spectra in all four examples, and their peak positions agree with the corresponding literature benchmarks. The spectra also resolve directional features, including the weaker normal IR response of monolayer MoS$_2$ and crossed Raman response of the Sb$_2$S$_3$ chain.

\FloatBarrier
%\vspace{6mm}

\subsection{Efficiency benchmark of ZStar}
\label{sec:unified_validation}

We timed separate BEC and phonon calculations and the Unified workflow for ten systems across four dimensionalities. Each pair uses the same geometry, functional, basis, SCF threshold, and nominal displacement of 0.02~bohr. The derivatives use the displacements measured from the saved structures. Table~\ref{tab:unified_efficiency} lists displacement counts, solver calls, and core-hours. Here $N_{\mathrm{NSCF}}$ counts PYATB calls.

\begin{table}[htbp]
\centering\small
\caption{Efficiency of the Separate and Unified workflows for BECs and zone-center force constants. $N_{\mathrm{BEC}}$ counts displaced structures. $N_{\mathrm{SCF}}$ includes the reference and the independent force calculations in Separate. $N_{\mathrm{NSCF}}$ counts PYATB calls, including the reference band check. Total cost is the sum of ABACUS and PYATB solver core-hours. Speedup is the Separate cost divided by the Unified cost.}
\label{tab:unified_efficiency}
\setlength{\tabcolsep}{2.5pt}
\begin{tabular*}{\textwidth}{@{\extracolsep{\fill}}lrrrrrrrrr@{}}
\toprule
 & \multicolumn{2}{c}{$N_{\mathrm{BEC}}$} & \multicolumn{2}{c}{$N_{\mathrm{SCF}}$} & \multicolumn{2}{c}{$N_{\mathrm{NSCF}}$} & \multicolumn{2}{c}{Total (core-h)} & \\
\cmidrule(lr){2-3}\cmidrule(lr){4-5}\cmidrule(lr){6-7}\cmidrule(lr){8-9}
System & Separate & Unified & Separate & Unified & Separate & Unified & Separate & Unified & Speedup \\
\midrule
cubic BaTiO$_3$ & 18 & 3 & 22 & 4 & 20 & 5 & 10.70 & 2.56 & 4.18 \\
3C-SiC & 12 & 2 & 15 & 3 & 14 & 4 & 13.16 & 3.78 & 3.48 \\
t-HfO$_2$ & 12 & 4 & 17 & 5 & 14 & 6 & 29.98 & 10.87 & 2.76 \\
$\alpha$-In$_2$Se$_3$ & 30 & 10 & 41 & 11 & 32 & 12 & 22.11 & 7.40 & 2.99 \\
hBN & 12 & 2 & 15 & 3 & 14 & 4 & 2.25 & 0.57 & 3.96 \\
MoS$_2$ & 12 & 3 & 16 & 4 & 14 & 5 & 7.22 & 3.01 & 2.40 \\
BN(9,0) & 60 & 56 & 117 & 57 & 62 & 58 & 270.41 & 125.04 & 2.16 \\
Sb$_2$S$_3$ & 30 & 20 & 51 & 21 & 32 & 22 & 41.67 & 18.95 & 2.20 \\
H$_2$O & 12 & 6 & 19 & 7 & 14 & 8 & 21.67 & 8.91 & 2.43 \\
CH$_4$ & 12 & 3 & 16 & 4 & 14 & 5 & 12.98 & 3.26 & 3.98 \\
\bottomrule
\end{tabular*}
\end{table}

Separate obtains BECs by central Cartesian differences and force constants from independent force calculations. Unified obtains both from the same symmetry-adapted displacements. Both workflows use equivalent 40-core nodes.

Table~\ref{tab:unified_spectra_efficiency} extends the comparison to IR and Raman spectra. Separate evaluates the Raman derivatives by central differences along all nonrigid modes, including Raman-inactive modes. Unified evaluates the derivatives with PYATB using the existing electronic matrices. The Raman comparison uses the same eigenvectors and frequencies for both workflows. Each IR spectrum uses the frequencies from its own workflow.

Unified reduces solver core-hours for BECs and zone-center phonons by a factor of up to 4.2. For BN(9,0), the number of BEC displacements falls from 60 to 56, and their force outputs replace 56 independent force calculations. Unified uses fewer core-hours for the combined IR and Raman calculation in all four cases. The largest speedup is 8.4-fold for CH$_4$.

\begin{center}
\begin{minipage}{\textwidth}
\centering
\captionsetup{hypcap=false,skip=4pt}
\captionof{table}{Efficiency of the Separate and Unified workflows for combined IR and nonresonant Raman spectra. The totals include the BEC and phonon calculations in Table~\ref{tab:unified_efficiency} and the Raman calculation. $N_{\mathrm R}$ counts additional normal-mode SCFs for Raman. $N_{\mathrm{SCF}}$ counts all SCFs and $N_{\mathrm{NSCF}}$ counts PYATB calls. Unified reuses the electronic matrices and adds no SCFs.}
\label{tab:unified_spectra_efficiency}
\footnotesize
\setlength{\tabcolsep}{2.5pt}
\begin{tabular*}{\textwidth}{@{\extracolsep{\fill}}lrrrrrrrrr@{}}
\toprule
 & \multicolumn{2}{c}{$N_{\mathrm R}$} & \multicolumn{2}{c}{$N_{\mathrm{SCF}}$} & \multicolumn{2}{c}{$N_{\mathrm{NSCF}}$} & \multicolumn{2}{c}{Total (core-h)} & \\
\cmidrule(lr){2-3}\cmidrule(lr){4-5}\cmidrule(lr){6-7}\cmidrule(lr){8-9}
System & Separate & Unified & Separate & Unified & Separate & Unified & Separate & Unified & Speedup \\
\midrule
t-HfO$_2$ (3D) & 30 & 0 & 47 & 5 & 44 & 11 & 87.39 & 15.48 & 5.65 \\
MoS$_2$ (2D) & 12 & 0 & 28 & 4 & 26 & 9 & 10.55 & 3.11 & 3.39 \\
Sb$_2$S$_3$ (1D) & 52 & 0 & 103 & 21 & 84 & 43 & 71.76 & 19.51 & 3.68 \\
CH$_4$ (0D) & 18 & 0 & 34 & 4 & 32 & 9 & 27.88 & 3.34 & 8.35 \\
\bottomrule
\end{tabular*}
\end{minipage}
\end{center}

\FloatBarrier
\section{Summary}
\label{sec:summary}

We have presented ZStar, an open-source toolkit that unifies the calculation of polarization, lattice dynamics, dielectric and piezoelectric responses, and vibrational spectra within one displacement workflow with ABACUS. From one symmetry-adapted set of displacements, the computed polarization and forces yield the Born effective charges (BECs) and zone-center force constants, and the electronic matrices obtained for the same configurations are reused to obtain Raman tensors. Berry-phase matching in periodic directions combined with real-space dipoles in nonperiodic directions extends the approach from bulk crystals (3D) to 2D and 1D materials and to molecules (0D).
Our benchmarks agree with published BECs, dielectric response, piezoelectric tensors, vibrational spectra and mode assignments. The workflow yields speedups of up to $4.2\times$ for BEC and zone-center phonon calculations and $8.4\times$ for combined infrared and Raman calculations. Beyond ABACUS, ZStar can post-process response data from VASP, CP2K, and Quantum ESPRESSO. Through its PYATB interface, it can also compute responses from machine-learned Hamiltonian models such as HamGNN, Hot-Ham, DeepH, and NextHAM. The first-principles data it generates can in turn be used to train field-aware machine-learned potentials. The \texttt{zstar-workflows} skill provides an agent-friendly interface that helps users get started smoothly. Coupling such workflows with machine-learned models should make it practical to screen large sets of structures for targeted dielectric, piezoelectric, or vibrational properties, and to carry the most promising candidates forward to device-scale studies.

\Needspace{5\baselineskip}
\section*{Data availability}

The source code, documentation, and examples are available at \url{https://github.com/xdzhu/ZStar}. The Python package is available at \url{https://pypi.org/project/zstar/}. The tetragonal HfO$_2$ reference data are available in \href{https://phonix-db.org/ver2026.03.28/data_1018001-1019000/mp-1018721.html}{Phonix entry mp-1018721} \cite{Ohnishi2026Phonix}. The single-chain Sb$_2$S$_3$ reference spectra are available in Ulian's \href{https://data.mendeley.com/datasets/6tntvw37tr/1}{Mendeley Data repository} \cite{Ulian2026Sb2S3Data}.

\section*{CRediT authorship contribution statement}

\textbf{Xudong Zhu}: Writing -- review and editing, Writing -- original draft, Conceptualization, Data curation, Formal analysis, Investigation, Methodology, Software, Validation, Visualization. \textbf{Junhong Li}: Investigation, Validation. \textbf{Gan Jin}: Methodology, Software. \textbf{Zheng Guan}: Software. \textbf{Meng Zhang}: Investigation. \textbf{Lixin He}: Writing -- review and editing, Supervision, Funding acquisition, Conceptualization.

\section*{Declaration of competing interest}

The authors declare that they have no known competing financial interests.

\section*{Acknowledgements}

This work was supported by the Anhui Provincial Science and Technology Breakthrough Project (Grant No. 202523o09050015), the Operating Funds of the Institute of Artificial Intelligence, Hefei Comprehensive National Science Center (Grant No. 22KT004), and the Innovation Program for Quantum Science and Technology (Grant No. 2021ZD0301200). Numerical computations were performed on the USTC HPC facilities and at the Hefei Advanced Computing Center.

\section*{Declaration of generative AI and AI-assisted technologies in the writing process}
During the preparation of this work, the authors used OpenAI Codex (GPT-6.1 Sol) to polish the text, visualize data, format tables, and assist with code debugging. The authors have manually verified the AI-generated content and take full responsibility for the accuracy, completeness, and originality of the entire manuscript.

\Needspace{12\baselineskip}

% Start of the appendix
\FloatBarrier
\begingroup
\let\bibfont\footnotesize
\bibliography{zstar}
\endgroup

\appendix
\numberwithin{table}{section}
\numberwithin{figure}{section}
\renewcommand{\thetable}{\Alph{section}.\arabic{table}}
\renewcommand{\thefigure}{\Alph{section}.\arabic{figure}}

\section{Molecular BEC tensors and HSE results}
\label{app:molecular_tensors}
The following charge-neutral tensors use rows for the dipole component and columns for atomic displacement, in units of $e$. This is the transpose of the displacement-first representation used in the corresponding output files. For H$_2$O, the molecular plane is $yz$, $x$ is its normal, and H$^+$ denotes the hydrogen at positive $y$ relative to O:
\begin{align}
 A_{\mathrm O}^{\mathrm{PBE}}&=\left(\begin{array}{@{}rrr@{}}-0.6375&0&0\\0&-0.4763&0\\0&0&-0.3278\end{array}\right), &
 A_{\mathrm H^+}^{\mathrm{PBE}}&=\left(\begin{array}{@{}rrr@{}}0.3188&0&0\\0&0.2382&0.0640\\0&0.0869&0.1639\end{array}\right),\\
 A_{\mathrm O}^{\mathrm{HSE}}&=\left(\begin{array}{@{}rrr@{}}-0.6683&0&0\\0&-0.4963&0\\0&0&-0.3545\end{array}\right), &
 A_{\mathrm H^+}^{\mathrm{HSE}}&=\left(\begin{array}{@{}rrr@{}}0.3341&0&0\\0&0.2482&0.0663\\0&0.0836&0.1773\end{array}\right).
\end{align}
For the other hydrogen, both $yz$ and $zy$ change sign. These two entries describe different dipole derivatives. For CH$_4$, choosing the hydrogen along the $(1,1,1)$ direction from carbon gives
\begin{equation}
 A_{\mathrm H}^{\mathrm{PBE}}=\left(\begin{array}{@{}rrr@{}}0.0052&-0.0670&-0.0670\\-0.0670&0.0052&-0.0670\\-0.0670&-0.0670&0.0052\end{array}\right),\qquad
 A_{\mathrm H}^{\mathrm{HSE}}=\left(\begin{array}{@{}rrr@{}}0.0037&-0.0656&-0.0656\\-0.0656&0.0037&-0.0656\\-0.0656&-0.0656&0.0037\end{array}\right).
\end{equation}
For this hydrogen site, the off-diagonal components are larger in magnitude than the mean diagonal charge.

The HSE calculations specify 25\% short-range exact exchange and a screening parameter of 0.11~bohr$^{-1}$ \cite{Krukau2006HSE}. They use a 100~Ry cutoff, 9-au numerical orbitals, a 20~\AA{} cubic cell, and an SCF threshold of $10^{-7}$. A separate exchange loop is enabled, with the exchange-screening thresholds given in the corresponding inputs. We integrate the ABACUS charge density at displacements of $\pm0.01$~\AA{} and take central differences. Raw tensors and pre-projection charge-sum residuals accompany the charge-neutral values. The efficiency benchmark uses the PBE molecular calculations.

\FloatBarrier

\Needspace{19\baselineskip}
\section{Electro-optic response and longitudinal Raman}
\label{app:electro_optic}

For AlAs, we calculate $\chi^{(2)}$ from the archived Quantum ESPRESSO dipoles at the smaller of the two field steps \cite{BastoneroMarzari2024,BastoneroVibroscopyData}. The $\chi^{(2)}_{xyz}$ result differs by about $0.001\%$ from the archive value (Table~\ref{tab:eo_benchmarks}). For AlP, both calculations use the Born charges, electronic permittivity, and susceptibility derivatives of the ABINIT tutorial \cite{ABINITNLOReference}. The longitudinal Raman response includes the macroscopic-field term. For the free-stress Pockels coefficient, we obtain the strain contribution from $p_{yz,4}$ evaluated at two relaxed-ion shear steps. The largest componentwise difference in the AlAs susceptibility is $1.2\times10^{-3}$~pm/V. For AlP, the displacement derivative alone gives a $yz$ Raman susceptibility of 0.00432.

\begin{center}
\begin{minipage}{0.76\textwidth}
\centering
\small
\captionsetup{hypcap=false}
\captionof{table}{Electro-optic and longitudinal Raman benchmarks. The AlP longitudinal modes propagate along [100], and the Raman susceptibility uses the ABINIT convention. The dash indicates that the tutorial provides no free-stress value for comparison.}
\label{tab:eo_benchmarks}
\begin{tabular}{llrr}
\toprule
Material & Quantity & ZStar & Reference \\
\midrule
AlAs & $\chi^{(2)}_{xyz}$ (pm/V) & 30.449 & 30.449 \\
AlP & Electronic part of $r_{yzx}$ (pm/V) & $-0.953$ & $-0.953$ \\
 & Ionic part of $r_{yzx}$ (pm/V) & $+0.536$ & $+0.536$ \\
 & $r_{yzx}$ at fixed strain (pm/V) & $-0.417$ & $-0.417$ \\
 & Strain contribution to $r_{yzx}$ (pm/V) & $-0.124$ & -- \\
 & $r_{yzx}$ at free stress (pm/V) & $-0.542$ & -- \\
 & Longitudinal frequency (cm$^{-1}$) & 534.69 & 534.69 \\
 & Longitudinal $yz$ Raman susceptibility & 0.00686 & 0.00686 \\
\bottomrule
\end{tabular}
\end{minipage}
\end{center}

\FloatBarrier
\section{Symmetry Classification of Zone-Center Optical Phonon Modes}
\label{app:point_group_phonons}

ZStar assigns $\Gamma$-point phonon symmetries and optical activity using the 32 crystallographic point groups. 
The optical-mode multiplicities of a crystal follow from its site symmetries and vibrational representation. IR-active species transform as components of a polar vector. First-order nonresonant Raman-active species transform as components of the symmetric polarizability tensor. In centrosymmetric groups these transformation rules require $u$ and $g$ parity, respectively. 
The selection rules from the Bilbao Crystallographic Server  POINT tables and standard vibrational analysis \cite{Aroyo2006BilbaoII,Rousseau1981NormalModes} are listed in Table~\ref{tab:point_group_phonons}.

\begin{center}
\begin{minipage}{\textwidth}
	\centering
	\small
	\setlength{\tabcolsep}{4pt}
	\renewcommand{\arraystretch}{1.05}
	\captionof{table}{Symmetry-allowed electric-dipole IR and first-order static nonresonant Raman species for zone-center optical phonons in the 32 crystallographic point groups.}
	\label{tab:point_group_phonons}
	\begin{tabularx}{\textwidth}{>{\hsize=0.4\hsize}X>{\hsize=1.4\hsize}X>{\hsize=0.5\hsize}X>{\hsize=0.5\hsize}X}
		\toprule
		\textbf{Point Group} & \textbf{Irreducible Representations} & \textbf{IR Active} & \textbf{Raman Active} \\
		\midrule
		$C_1$ & $A$ & $A$ & $A$ \\
		$C_i$ & $A_g$, $A_u$ & $A_u$ & $A_g$ \\
		$C_2$ & $A$, $B$ & $A$, $B$ & $A$, $B$ \\
		$C_s$ & $A'$, $A''$ & $A'$, $A''$ & $A'$, $A''$ \\
		$C_{2h}$ & $A_g$, $B_g$, $A_u$, $B_u$ & $A_u$, $B_u$ & $A_g$, $B_g$ \\
		$D_2$ & $A$, $B_1$, $B_2$, $B_3$ & $B_1$, $B_2$, $B_3$ & $A$, $B_1$, $B_2$, $B_3$ \\
		$C_{2v}$ & $A_1$, $A_2$, $B_1$, $B_2$ & $A_1$, $B_1$, $B_2$ & $A_1$, $A_2$, $B_1$, $B_2$ \\
		$D_{2h}$ & $A_g$, $B_{1g}$, $B_{2g}$, $B_{3g}$, $A_u$, $B_{1u}$, $B_{2u}$, $B_{3u}$ & $B_{1u}$, $B_{2u}$, $B_{3u}$ & $A_g$, $B_{1g}$, $B_{2g}$, $B_{3g}$ \\
		$C_4$ & $A$, $B$, $E$ & $A$, $E$ & $A$, $B$, $E$ \\
		$S_4$ & $A$, $B$, $E$ & $B$, $E$ & $A$, $B$, $E$ \\
		$C_{4h}$ & $A_g$, $B_g$, $E_g$, $A_u$, $B_u$, $E_u$ & $A_u$, $E_u$ & $A_g$, $B_g$, $E_g$ \\
		$D_4$ & $A_1$, $A_2$, $B_1$, $B_2$, $E$ & $A_2$, $E$ & $A_1$, $B_1$, $B_2$, $E$ \\
		$C_{4v}$ & $A_1$, $A_2$, $B_1$, $B_2$, $E$ & $A_1$, $E$ & $A_1$, $B_1$, $B_2$, $E$ \\
		$D_{2d}$ & $A_1$, $A_2$, $B_1$, $B_2$, $E$ & $B_2$, $E$ & $A_1$, $B_1$, $B_2$, $E$ \\
		$D_{4h}$ & $A_{1g}$, $A_{2g}$, $B_{1g}$, $B_{2g}$, $E_g$, $A_{1u}$, $A_{2u}$, $B_{1u}$, $B_{2u}$, $E_u$ & $A_{2u}$, $E_u$ & $A_{1g}$, $B_{1g}$, $B_{2g}$, $E_g$ \\
		$C_3$ & $A$, $E$ & $A$, $E$ & $A$, $E$ \\
		$C_{3i}$ & $A_g$, $E_g$, $A_u$, $E_u$ & $A_u$, $E_u$ & $A_g$, $E_g$ \\
		$D_3$ & $A_1$, $A_2$, $E$ & $A_2$, $E$ & $A_1$, $E$ \\
		$C_{3v}$ & $A_1$, $A_2$, $E$ & $A_1$, $E$ & $A_1$, $E$ \\
		$D_{3d}$ & $A_{1g}$, $A_{2g}$, $E_g$, $A_{1u}$, $A_{2u}$, $E_u$ & $A_{2u}$, $E_u$ & $A_{1g}$, $E_g$ \\
		$C_6$ & $A$, $B$, $E_1$, $E_2$ & $A$, $E_1$ & $A$, $E_1$, $E_2$ \\
		$C_{3h}$ & $A'$, $E'$, $A''$, $E''$ & $A''$, $E'$ & $A'$, $E'$, $E''$ \\
		$C_{6h}$ & $A_g$, $B_g$, $E_{1g}$, $E_{2g}$, $A_u$, $B_u$, $E_{1u}$, $E_{2u}$ & $A_u$, $E_{1u}$ & $A_g$, $E_{1g}$, $E_{2g}$ \\
		$D_6$ & $A_1$, $A_2$, $B_1$, $B_2$, $E_1$, $E_2$ & $A_2$, $E_1$ & $A_1$, $E_1$, $E_2$ \\
		$C_{6v}$ & $A_1$, $A_2$, $B_1$, $B_2$, $E_1$, $E_2$ & $A_1$, $E_1$ & $A_1$, $E_1$, $E_2$ \\
		$D_{3h}$ & $A_1'$, $A_2'$, $E'$, $A_1''$, $A_2''$, $E''$ & $A_2''$, $E'$ & $A_1'$, $E'$, $E''$ \\
		$D_{6h}$ & $A_{1g}$, $A_{2g}$, $B_{1g}$, $B_{2g}$, $E_{1g}$, $E_{2g}$, $A_{1u}$, $A_{2u}$, $B_{1u}$, $B_{2u}$, $E_{1u}$, $E_{2u}$ & $A_{2u}$, $E_{1u}$ & $A_{1g}$, $E_{1g}$, $E_{2g}$ \\
		$T$ & $A$, $E$, $T$ & $T$ & $A$, $E$, $T$ \\
		$T_h$ & $A_g$, $E_g$, $T_g$, $A_u$, $E_u$, $T_u$ & $T_u$ & $A_g$, $E_g$, $T_g$ \\
		$O$ & $A_1$, $A_2$, $E$, $T_1$, $T_2$ & $T_1$ & $A_1$, $E$, $T_2$ \\
		$T_d$ & $A_1$, $A_2$, $E$, $T_1$, $T_2$ & $T_2$ & $A_1$, $E$, $T_2$ \\
		$O_h$ & $A_{1g}$, $A_{2g}$, $E_g$, $T_{1g}$, $T_{2g}$, $A_{1u}$, $A_{2u}$, $E_u$, $T_{1u}$, $T_{2u}$ & $T_{1u}$ & $A_{1g}$, $E_g$, $T_{2g}$ \\
		\bottomrule
	\end{tabularx}
\end{minipage}
\end{center}

\clearpage
\section{Polarization in two-dimensional materials}
\label{app:electrostatic_potential}
\setcounter{figure}{0}

\begin{figure*}[!htbp]
	\centering
\includegraphics[width=0.82\textwidth]{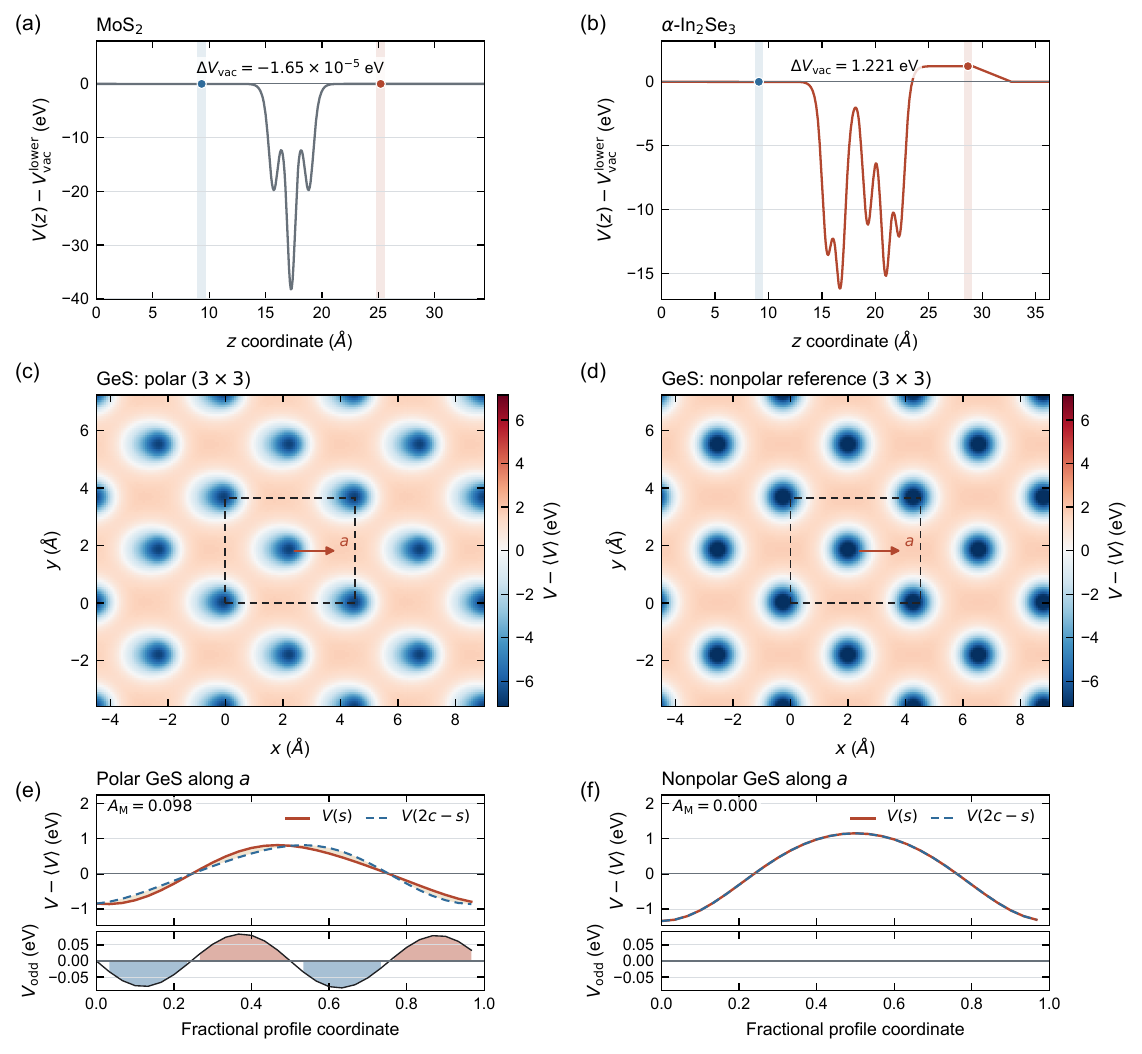}
	\caption{Electrostatic-potential diagnostics. (a,b) Plane-averaged slab-normal potentials of MoS$_2$ and $\alpha$-In$_2$Se$_3$; shaded windows define the vacuum plateaus. (c,d) Mean-centered in-plane potential maps of polar GeS and its fixed-ion nonpolar reference, using the same color scale. The maps show a $3\times3$ tiling of the computed cell. Dashed boxes mark the unitcell. (e,f) Potentials along $a$ and their optimally aligned mirror images, with mirror-odd components below. Profile and residual scales are shared between the two structures.}
	\label{fig:potential_examples}
\end{figure*}

Out-of-plane polarization in an isolated two-dimensional material produces a dipole normal to the sheet and a potential step between the two vacuum regions. For a slab normal to $z$, we calculate the planar average $\overline{V}(z)=A_{\parallel}^{-1}\int V(x,y,z)\,dx\,dy$. The vacuum levels are obtained from the plateau regions away from the dipole-correction discontinuity. Figure~\ref{fig:potential_examples}(a,b) shows the results for nonpolar MoS$_2$ and polar $\alpha$-In$_2$Se$_3$. Both vacuum plateaus of nonpolar MoS$_2$ round to 2.967~eV [Fig.~\ref{fig:potential_examples}(a)], with a residual difference of nearly zero. Polar $\alpha$-In$_2$Se$_3$ has plateaus of 3.064 and 4.285~eV [Fig.~\ref{fig:potential_examples}(b)], giving $\Delta V_{\mathrm{vac}}=1.221$~eV.

For GeS, we compare the polar $Pmn2_1$ structure with a fixed-cell $Pmmn$ reference. Following Fei \textit{et al.}\ \cite{Fei2016GeS}, we remove the relative Ge/S offset along the armchair direction and restore equal-species inversion pairs. Figure~\ref{fig:potential_examples}(c,d) shows the two potential maps with $3\times3$ cells. Along $a$, we fit the reflection center $c$ on each one-period profile by minimizing
\begin{equation}
 A_{\mathrm M}=\min_c\frac{\|\widetilde V(s)-\widetilde V(2c-s)\|_2}{2\|\widetilde V(s)\|_2},
 \qquad \widetilde V(s)=V(s)-\langle V\rangle.
\end{equation}
The polar GeS profile has $A_{\mathrm M}=0.098$ and a mirror-odd RMS amplitude of 0.058~eV [Fig.~\ref{fig:potential_examples}(e)]. The nonpolar profile gives zero.

\FloatBarrier

\section{Example Input Files}
\label{app:input_files}

\subsection{Input file for ABACUS SCF calculation in a BEC workflow}

\begin{lstlisting}[keepspaces=true]
INPUT_PARAMETERS
ecutwfc                     100
calculation                 scf
ks_solver                   genelpa
basis_type                  lcao
nspin                       1
scf_nmax                    200
scf_thr                     1e-7
init_chg                    auto
stru_file                   STRU
out_mat_hs2                 1
out_mat_r                   1
out_chg                     1
cal_force                   1
kspacing                    0.1
dft_functional              pbesol
\end{lstlisting}

\medskip
\subsection{Input file for PYATB polarization in a BEC workflow}

\begin{lstlisting}[keepspaces=true]
INPUT_PARAMETERS
{
    nspin                   1
    package                 ABACUS
    fermi_energy            20.301527396
    fermi_energy_unit       eV
    HR_route                /path/to/OUT.ABACUS/data-HR-sparse_SPIN0.csr
    SR_route                /path/to/OUT.ABACUS/data-SR-sparse_SPIN0.csr
    rR_route                /path/to/OUT.ABACUS/data-rR-sparse.csr
    HR_unit                 Ry
    rR_unit                 Bohr
}

LATTICE
{
    lattice_constant        1.0
    lattice_constant_unit   Angstrom
    lattice_vector
    3.56170284  0.00000000  0.00000000
    0.00000000  3.56170284  0.00000000
    0.00000000  0.00000000  5.16000037
}

POLARIZATION
{
    occ_band                38
    nk1                     22
    nk2                     22
    nk3                     15
    atom_type               2
    stru_file               STRU
    valence_e               26 6
}
\end{lstlisting}

\end{document}